\documentclass[11pt,aps,preview,amsmath,amssymb]{revtex4}

\usepackage{graphicx}
\usepackage{url}
\usepackage{multirow}
\usepackage{setspace}
\usepackage{tabularx}
\usepackage{color}

\usepackage[ pdftex, plainpages = false, pdfpagelabels,
         pdfpagelayout = useoutlines,
         bookmarks,
         bookmarksopen = true,
         bookmarksnumbered = true,
         breaklinks = true,
         linktocpage=all,
         pagebackref=false,
         colorlinks = true,
         linkcolor = BrickRed,
         urlcolor = blue,
         citecolor = BrickRed,
         anchorcolor = green,
         hyperindex = true,
         hyperfigures
         ]{hyperref}

\usepackage{cancel}
\usepackage[dvipsnames,tables]{xcolor}

\makeatletter
\g@addto@macro{\UrlBreaks}{\UrlOrds}
\makeatother

\begin{document}
\title{A Possible Link Between Pyriproxyfen and Microcephaly}
\author{Yaneer Bar-Yam$^*$, H. Frederik Nijhout$^\dag$, Raphael Parens$^*$, Felipe Costa$^*$, and Alfredo J. Morales$^*$}
\affiliation{ $^*$New England Complex Systems Institute, 210 Broadway Suite 101, Cambridge MA 02139,  $^\dag$Department of Biology, Duke University, Durham, NC 27708} 
\date{\today}
\begin{abstract}
The Zika virus has been the primary suspect in the large increase in incidence of microcephaly in 2015-6 in Brazil. However its role is not confirmed despite individual cases in which viral infections were found in neural tissue. Recently, the disparity between the incidences in different geographic locations has led to questions about the virus's role. Here we consider the alternative possibility that the use of the insecticide pyriproxyfen for control of mosquito populations in Brazilian drinking water is the primary cause.
Pyriproxifen is a juvenile hormone analog which has been shown to correspond in mammals to a number of fat soluble regulatory molecules including retinoic acid, a metabolite of vitamin A, with which it has cross-reactivity and whose application during development has been shown to cause microcephaly. 
Methoprene, another juvenile hormone analog that was approved as an insecticide based upon tests performed in the 1970s, has metabolites that bind to the mammalian retinoid X receptor, and has been shown to cause developmental disorders in mammals. 
Isotretinoin is another example of a retinoid causing microcephaly in human babies via maternal exposure and activation of the retinoid X receptor in developing fetuses.
Moreover, tests of pyriproxyfen by the manufacturer, Sumitomo, widely quoted as giving no evidence for developmental toxicity, actually found some evidence for such an effect, including low brain mass and arhinencephaly---incomplete formation of the anterior cerebral hemispheres---in exposed rat pups. Finally, the pyriproxyfen use in Brazil is unprecedented---it has never before been applied to a water supply on such a scale. Claims that it is not being used in Recife, the epicenter of microcephaly cases, do not distinguish the metropolitan area of Recife, where it is widely used, and the municipality, and have not been adequately confirmed. Given this combination of information about molecular mechanisms and toxicological evidence, we strongly recommend that the use of pyriproxyfen in Brazil be suspended until the potential causal link to microcephaly is investigated further. 
\end{abstract}
\maketitle


\section{Overview}
In 2015 and early 2016 over 6,000 suspected \cite{MdS2016_2} cases of microcephaly and other neurodevelomental disorders were reported in Brazil, primarily in the northeast, a dramatic increase over expected numbers. By September 2016, over 2,000 were confirmed \cite{WHO}. The cause of these developmental disorders has been widely attributed to a Zika virus epidemic first detected in May, 2015 \cite{WHO2015}. While evidence exists, particularly in neurological infections of a few microcephalic cases \cite{Mlakar2016,Tang2016}, only $12-16\%$ of confirmed microcephaly cases have also been confirmed as having Zika infections \cite{MdS2016_2,bul7272016}, preventing health authorities from determining a conclusive causal link \cite{schuler2016,Parens2016}. The strongest potential counter evidence is present in the geographic distribution of cases in Brazil and a significantly lower number of government reported cases in Colombia, though questions about the way cases are reported prevent definite conclusions. Reports suggest that areas outside of northeast Brazil do not have microcephaly cases commensurate with the prevalence of Zika infections \cite{naturebutler}.

The first cases of Zika in Colombia reportedly occurred in September 2015 \cite{WHO2015}.  During the first 12 full weeks of 2016, 34 cases of microcephaly were identified  \cite{Poho8April2016,WHOApril72016}, compared to 32 cases expected based upon 140 annual background cases. Subsequently, cases with evidence of both Zika and microcephaly were analyzed \cite{August1}. Background cases of microcephaly that are coincidentally present in pregnancies that have Zika infections account for 11 cases until Nov. 1, 2016. An additional 45 cases for a total of 56 cases can be attributed to Zika infections at the end of the first trimester. These cases are quite limited in number, and much lower than expectations \cite{Cobb2016}. In order for a consistent conclusion that Zika is the primary cause of microcephaly, the reported cases of Zika in Colombia and Brazil would have to be in the same ratio; with 90,000 reported Zika cases in Colombia and 200,000 in Brazil, they are widely discrepant. Reports of the number of Zika infections are not reliable and there are differences in methods of microcephaly counting, making definite conclusions difficult. However, the two orders of magnitude separating the number of cases in Brazil and Colombia certainly raise questions about whether Zika is the dominant cause or whether there are other causes or cofactors that are increasing the incidence of microcephaly in Brazil. There continue to be many more cases in Northeast Brazil at a rate of approximately 100 per month. Other countries report very few cases \cite{WHO,sitrep}.

A group of physicians in Argentina and Brazil have suggested that widespread use of the pesticide pyriproxifen \cite{Pubchem} to reduce mosquito populations because of a dengue epidemic in 2014 may be the cause of microcephaly \cite{Physicians2016}. This suggestion has been challenged both by authorities based upon a claim of lack of evidence \cite{Costa2016,Debunk} and by skeptics of conspiracy theories \cite{Gorski2016}. We have previously reviewed the primary evidence \cite{Parens2016}, including evidence in rat toxicology studies performed by its manufacturer Sumitomo \cite{Saegusa1988}, concluding that pyriproxifen is a credible candidate so that further inquiry is warranted. The Swedish Toxicology Research Center has published a review of toxicological information and potential molecular mechanisms associated to thyroid function and concluded that further studies are warranted \cite{swetox}. Moreover, arguments that toxicology studies are sufficient to rule it out as a cause would not rule it out as a cofactor. 

Here we discuss the molecular mechanisms associated with pyriproxyfen in insect and mammalian development and summarize toxicological studies and insecticide use in Brazil \cite{Evans2016}. We also review available data on Zika infections. Of particular note is that pyriproxyfen, a biochemical analog of juvenile hormone in insects, is cross-reactive with the retinoic acid / vitamin A regulatory system of mammals, and that exposure to retinoic acid has been shown to cause microcephaly \cite{Encyclopedia}. Existing toxicological studies and experience with public use do not provide evidence to counter this causal chain. Toxicological studies, if anything, provide evidence for neurodevelopmental toxicity at a level consistent with that found for microcephaly in Brazil. Pyriproxyfen's use in drinking water is unprecedented. The few trials using pyriproxyfen in the water supply did not investigate nor were they sufficient to observe a role in causing birth defects at the level seen in Brazil. Our review of the currently available information points to adequate reasons to suspect this insecticide as a developmental disruptor causing microcephaly. It also provides reasons to question the adequacy of the testing of pyriproxyfen's toxicology and its use in water supplies.

This paper is organized as follows. In Section II, we describe the potential causes of microcephaly. In Section III, we discuss the molecular mechanisms of pyriproxyfen. In Section IV, we review animal toxicological studies conducted in laboratory settings. In Section V, we describe pyriproxyfen usage in drinking water supplies in Brazil. In Section VI, we discuss evidence for and against the causal relationship between Zika and microcephaly and analyze data on Zika and microcephaly from Brazil and Colombia. In Section VII, we summarize the evidence and suggest policy changes based upon the available data.

\section{Potential causes and timeline}

The World Health Organization (WHO) identifies the most common causes of microcephaly as: infections, exposure to toxic chemicals, genetic abnormalities, and severe malnutrition while in the womb. The first includes toxoplasmosis, which is caused by parasites in undercooked meat, rubella, herpes, syphilis, cytomegalovirus and HIV. The second includes exposure to heavy metals, such as arsenic and mercury, as well as alcohol, smoking and radiation exposure. The third may include Down syndrome \cite{WHO2016}. While the first category points to the possibility of infectious causes, the second points to the possibility of chemical ones. The existence of multiple causes of microcephaly suggests many potential origins of a general developmental disruption. Hence, it is possible that more than one source of microcephaly is present in Brazil.

The suspected association of increased microcephaly cases with Zika and/or pyriproxyfen relies first on an increase in the cause with the increase in number of cases. The outbreak of the former was recognized in May 2015, and the use of the latter in the fourth quarter of 2014, with the incidence of microcephaly cases beginning in October of 2015. Thus both Zika and pyriproxyfen satisfy the first criterion for a cause. The precise dates of the increase in cases are believed to be uncertain because of potential problems with underreporting prior to the medical alert and overreporting afterwards \cite{schuler2016}. 

\section{Pyriproxyfen molecular mechanisms}
		
Juvenile hormone and retinoic acid are both lipid-soluble terpenoids.  They act as signaling molecules that control a broad diversity of embryonic and postembryonic developmental processes in insects and vertebrates, respectively.  Juvenile hormone is best known for its role in the control of metamorphosis and reproduction in insects \cite{Riddiford2012, Wheeler2003}, whereas retinoic acid is involved in the development of the nervous system in vertebrates \cite{Rhinn2012}.  

These two classes of hormone-like molecules share some molecular similarities and are capable of some degree of cross-reactivity.  For instance, retinoic acid is known to mimic some of the effects of juvenile hormone when injected into insects \cite{Nemec1993}, and juvenile hormone and its analogs are known to bind to the vertebrate retinoic acid receptor \cite{Palli1991, Jones1995, Harmon1995}.  It is possible therefore that pyriproxyfen, a powerful juvenile hormone analog \cite{Dhadialla1998}, can bind to the retinoic acid receptor.  When it does so it could either activate the receptor at inappropriate times in development, or act as a blocker that prevents the normal retinoic acid from binding to the receptor when needed.  The retinoic acid receptor normally turns on gene expression in development, so either an inappropriate activation or an inhibition at a critical time could be expected to lead to developmental abnormalities. 

More specifically, another juvenile hormone analog that has been approved as an insecticide, methoprene, has also been shown to have metabolites that bind to the mammalian retinoid X receptor and has been shown to cause developmental disorders in mammals \cite{Harmon1995,Unsworth1974}. 

Isotretinoin is a retinoid that is widely used in medicine but is contraindicated in women who are pregnant or might become pregnant. It causes microcephaly in human babies via maternal exposure and activation of the retinoid X receptor in developing fetuses \cite{Stem1989,Irving1986}.

The impact of retinoids on abnormal development (teratogenesis) has been demonstrated to be sensitive to genotype and developmental stage of exposure and to result in death, malformation, growth retardation,
and/or functional disorder \cite{Collins1999}. The juvenile hormone itself and different juvenile hormone analogs have different binding properties to mammalian retinoic acid receptors, which could therefore produce different effects, including teratogenicity. These effects remain poorly understood \cite{Wheeler2003,Flat2006}.

\section{Toxicological Studies}

\subsection{Overview of Toxicological Studies}

The potential link between pyriproxyfen and microcephaly has been challenged based upon the existence of toxicological studies by the manufacturer Sumitomo. However, we have reviewed their studies and find them largely restricted to analysis of the impact on adult animals. In the limited experiments on developmental toxicity, the ability to identify a link to microcephaly is weak based upon the specific tests that have been performed. Rather than measurements, in most tests visual macroscopic observation of rat and rabbit fetuses or pups/kits were used \cite{sumitomo}, for which standards of microcephaly determination may not be sufficiently well established or comparable across species. In the most relevant experiment there is a reported test of brain weight and neurodevelopmental disorders in rat pups, and in that study, there is actual evidence for microcephaly. Sumitomo has argued that the evidence for microcephaly should not be considered because of its dose dependence, but this argument is not consistent with the statistical nature of the study, as the dependence on dosage is not statistically verified. Further, it is important to recognize that the study design adopts an ``innocent until proven guilty" assumption as the null hypothesis is the absence of toxicity. It is the presence of toxicity that must pass a statistical significance test, which is not the same as the proof of its absence.   

Specifically, the most relevant study for a determination of neurodevelopmental toxicity \cite{Saegusa1988} considered brain and behavioral effects of rat pups exposed to pyriproxyfen during days 7 to 17 of gestation, which lasts 21 days. 
The experimental group of 36 pregnant rats in each of four test groups was fed dosage levels of 0, 100, 300 and 1000 mg/kg/day. The pups were checked for physiological deformations and organs weighed. 

From each dosage level, litters of pups were obtained. While the target was to obtain 99 pups in each group, only 78 pups were obtained in the 1000 mg/kg group due to adult deaths. For 99 pups in the 100 mg/kg and 78 pups of the 1000 mg/kg dosage groups no relevant developmental disorders were reported. Of the 99 pups in the 300 mg/kg dosage 1 (1\%) had Arhinencephaly and 1 (1\%) had Thyroid hypoplasia. The former would be consistent with concerns about neurodevelopmental disorders of the type of microcephaly as well as observations that a variety of neurological and other developmental disorders have been found associated with the epidemic in Brazil \cite{Brasil2016}.

Of the resulting offspring, a number of pups were kept alive for emotional/mental testing at 4 and 6 weeks of age and their brains were subsequently weighed at 8 weeks. The report then gives values for group sizes of 13, 12, 11, and 10 for both male and female in the control and the three dosage levels, respectively. One of the groups, the males of the 300 mg/kg group, had statistically significant lower brain weight at 8 weeks. Male pups at other dosage levels were also lower in relative brain mass compared to the control.

These tests provide evidence that microcephaly may be an outcome of the application of pyriproxyfen to rats and other mammals. As discussed in the following section, the measures of low brain mass and incidence of Arhinencephaly were dismissed by Sumitomo based upon the assumption of dose dependence---higher dosages should lead to larger toxicological effects, otherwise the effects must not be toxicological. The conventional toxicology assumptions of dose dependence infer from the absence of similar observations in the 1,000 mg/kg dosage group that the observations in the 300 mg/kg group are not relevant to the toxicology of pyriproxyfen \cite{sumitomo}. 

This assumption, however was not correctly applied due to the statistical nature of the experiment: Statistical variation may lead to lower effects by random chance, which may mask increasing effects. Tests of toxicological effects must demonstrate that the direction of change is counter to a toxicological effect to an observed level of statistical certainty. 

This is also a concern for individual events such as the case of Arhinencephaly. Rare events should not be assumed to be background effects when those effects may also be caused by toxicological effects that are rare. Specific events may be caused by genetic variability and physiological regulatory sensitivity. Instead, the outcomes might be interpreted as an estimated probability of incidence of 1 in 66, the total number of pups in the observed exposure groups. Considering the differences between humans and rats, the values for humans may be higher or lower than this. Note that the incidence of microcephaly in Brazil is estimated to be on the order of a few percent \cite{MdS2016_2}. If the rate in rats and human beings is comparable, then it would be expected that only one event would happen among the entire rat pup population of the experiment. The need for careful studies is apparent in that teratogenicity of a retinoid varies widely across mammalian species, and this variation itself varies across different retinoids \cite{Wilhite1986, Irving1986, Tzimas1994, Nau1993, Eckhoff1997}. Regulatory systems are remarkably sensitive to small molecular densities. Given the low incidence of human cases of microcephaly in affected populations, experiments that are designed to test for microcephaly in mammals, at the very least, should be designed to identify incidence at this level. Moreover, they should be unambiguously capable of detecting microcephaly. 

Finally, there are multiple drugs that passed conventional regulatory animal testing and are now known to be linked to microcephaly, including phenytoin and methotrexate \cite{Homes2001, Hyoun2012, swetox}, so assurances based solely upon conventional regulatory testing are inadequate in this case as well.

\subsection{Toxicological tests are framed to require proof of toxicity rather than safety}

Given the evidence for low brain mass, it is important to inquire why Sumitomo researchers and regulatory authorities dismiss the risks of pyriproxyfen. More importantly, why they consider the issue to be settled as to the absence of risk from pyriproxyfen. 

As indicated above, this is particularly surprising given the underlying framing of the toxicological studies. The statistical test of toxicity takes as its null hypothesis that there is no toxicity. The use of statistical significance for toxicity assumes that proof of toxicity is required, not of safety. This is counter to the premise that the toxicity study provides evidence of a lack of toxicity. Absence of proof of toxicity is surely not proof of the absence of toxicity. Given this framing, it should not be surprising that many cases where toxicological tests do not find toxicity result in the approval of substances that are later found to be toxic.

In particular, the null hypothesis of the statistical study is that there is no difference between the control and the sample that has received pyriproxyfen. The type of test that is done is similar to that of a medicine which requires proof of positive effect, satisfying a  statistical significance threshold, in order to be administered. In the case of toxicity, the absence of statistical significance is not a demonstration of the lack of toxicity, but rather a lack of proof of toxicity. Thus the study is not considering whether pyriproxyfen is safe, it is asking whether there is proof of toxicity beyond a statistical uncertainty.  

This logical framework also leads to a circumstance where low statistical power because of small sample numbers decreases the ability of the study to identify actual toxicity. In such a case, the absence of statistical significance may directly follow from poor statistics. In a commercial context where regulatory approval enables sales and profits, there is a financial incentive to use small samples that minimize the statistical power of the test, because a limited test does not show toxicity even when it is present. 

In this case, however, there is actually a significant result for one of the trials. The average effect is large enough even in a context where the standard error is large because of the small sample used. 

However, the report, and subsequent government regulatory approval, dismisses the result because there is not a larger effect at higher dosage. This dose dependent assumption is standard in toxicology. However, the dose dependence, which is only of two data points, itself is not subject to a statistical test. The trend is considered to be true even though it is not valid within statistical uncertainty---the opposite trend of an increasing impact could occur given statistical variation between the two dosage levels. Smaller brain mass relative to the control is present in the higher dosage case. While less severe on average, the difference between the lower and higher dosage level is well within a standard error, and therefore is within observational uncertainty. Thus it should not be used to make decisions about whether there is toxicity or not. 

The study does not provide evidence for an absence of toxicity. It assumes a null effect, has poor statistical power, actually has some evidence for toxicity, and that evidence is dismissed based upon incorrect use of a trend analysis that statistics does not support. 

\subsection{Specifics of Sumitomo studies of pyriproxyfen developmental toxicity}

The toxicity experiments summarized above were performed by the producer of pyriproxyfen, Sumitomo, in 1988. The most relevant of these studies is a rat teratology study by administration of pyriproxyfen to pregnant rats (dams) by gavage on days 7-17 after pregnancy \cite{Saegusa1988}. A second study was performed on rabbit developmental toxicity by administration on days 6-18 of pregnancy \cite{Wilkinson1994}. In this section we provide additional detail about these studies. 

\subsubsection{Low brain mass observations (rat teratogenicity study)}

\begin{figure*}[tb]
\begin{center}
\includegraphics[scale=.65]{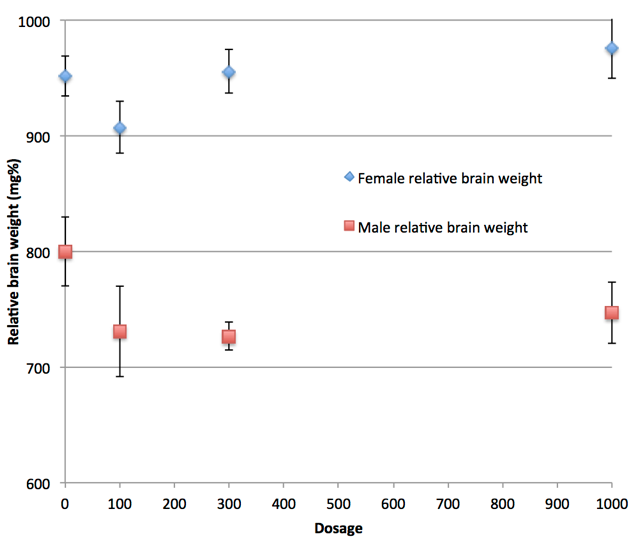}
\caption{Brain weight of male rat pups relative to body weight on postnatal day 56. Ranges are standard error of the mean. One value is statistically signficant (300 mg/kg). The importance of this value is dismissed by Sumitomo (and we can infer by regulatory authorities who approved the use of pyriproxyfen in drinking water). The reason for this dismissal is that the 1000 mg/kg dosage value is not statistically significant and it is assumed that if pyriproxyfen is the cause of the low brain mass then the higher dosage level would also give a statistically significant result. Note, however, that the 1,000 mg/kg value (as well as the 100 mg/kg value) are low relative to the control and their values are not statistically different from the 300 mg/kg value. Remarkably, this failure of proper statistical reasoning appears to be the reason for regulatory approval.  
\label{datafigure}}
\end{center}
\end{figure*}

In the rat teratogenicity study, four groups were administered different amounts of pyriproxyfen, 0 mg/kg (control), 100mg/kg, 300 mg/kg, and 1000 mg/kg. Of particular importance to our analysis is the observation of a statistically low relative brain weight among males in the 300mg/kg group (see Table \ref{weights} and Fig. \ref{datafigure}). 

\begin{table}[t]
\begin{tabularx}{\textwidth}{|X|X|X|X|X|X|}
\hline
Gender & Measure & Control  & 100 mg/kg  & 300 mg/kg  & 1,000 mg/kg  \\
 &  & (13 pups)$^\ddag$ & (12 pups)$^\ddag$ & (11 pups)$^\ddag$ &  (10 pups)$^\ddag$ \\
\hline
Male & Body weight (g) &$232.2\pm30.2$ & $252.0\pm33.8$ & $253.4\pm20.1$&$257.8\pm40.2$\\ \hline
Male & Brain weight (g) & $1.8\pm0.101$ &$1.817\pm0.079$&$1.859\pm0.041$ &$1.809\pm.097$ \\ \hline
Male & Brain relative weight (mg$\%^\dag$) & $799.9\pm103$& $730.8\pm130.9$ & $726.6\pm 38.6^*$ & $746.9 \pm 80.6$ \\ \hline
Female & Body weight (g) & $171.1\pm11.1$ & $182.1\pm16.3$ & $170.6\pm6.2$ & $168.5\pm13$ \\ \hline       
Female & Brain weight (g) & $1.711\pm.091$ & $1.737\pm.077$ & $1.771\pm.041^*$ & $1.708 \pm .082$ \\ \hline
Female & Brain relative weight (mg$\%^\dag$) & $951.7\pm60.4$& $907.4\pm75.9$ & $955.6 \pm 59.1$ & $976.4\pm 80$ \\ \hline

\end{tabularx}
\caption{Body, brain and brain relative weights (mean and standard deviation) for pups separated into groups by gender and dosage  \cite{Saegusa1988}. $^*$ Statistically significant values. $^\dag$ Notation used in the study for ratio of brain weight in mg per 100 g body weight.  $^\ddag$ Progressively smaller numbers of pups not explained. } \label{weights}
\end{table}

Low brain weight measurements may be indicative of microcephaly. However, according to communications from Sumitomo, results were interpreted according to standard assumptions of dose dependence. According to this assumption deviations at lower dosages are not considered if they don't also occur at higher dosage levels. Thus, the assumptions of dose dependence dismiss the findings of lower relative brain weights in lower dose groups (300 mg/kg group), as brain weights is observed in higher dose groups (1000 mg/kg) are not statistically significant \cite{Saegusa1988}.

The assumption of dose dependence is a mathematical assumption about the distribution of observations resulting from variables $X$ that cause variations in the measured quantity unrelated to the application of pyriproxyfen, and $Y$ that cause variations due to pyriproxyfen. It is assumed that the observable (relative brain weight)
\begin{equation}
W(p)  =  \sum_{X,Y} W(X,Y) P(X,Y;p),
\label{Eq_av}
\end{equation}
is a monotonically varying function of $p$. However, this assumption is then applied to the observed value, which is not given by the actual average overall probabilities but by a sample:
\begin{equation}
W_o(p_j)  =  \sum_{X_i,Y_i} W(X_i,Y_i),
\label{Eq_2}
\end{equation}
where the sum is over the particular cases that arose in the sampling of the experiment for dosage $p_j$. The difference between observed and expected outcomes $W_o(p_j)-W(p)$ has distribution of values. The conclusion of monotonic dose dependence for a teratogenic effect would only be clearly valid if the standard deviation of this distribution is small compared to the difference of values observed. More precisely, a null hypothesis of a teratogenic effect should be falsified by the observational study. However, the standard error of the observations is larger than the difference between those observations and therefore does not support this conclusion.

More specifically, all three groups administered pyriproxyfen are low in relative brain mass compared to the control and differ from each other by statistically insignificant amounts.The 300 mg/kg group has a value of 726.6, compared to the control of 799.9, a difference of 73.3. The 100 mg/kg group has a value of 730.8, a difference relative to the control of 69.1, only 4.2 smaller. The 1000 mg/kg group value is 746.9 which is 53 less than the control. While the latter two are not statistically significant according to the analysis, the 100 mg/kg case is not far from statistical significance, and taken together, the set of three results across dosage levels provide evidence that pyriproxyfen causes low brain weight.

We note that a relatively high number of pregnant rats in the 1000 mg/kg died before pregnancy (12 of 42 pregnant rats) \cite{Saegusa1988}. Those who survived the test are not an unbiased sample (survivor bias), perhaps eliminating cases of low relative brain weight among pups in this class. This would follow from a circumstance in which those rat pups who were more susceptible to pyriproxyfen leading to low brain mass, would also be the ones who, at increased dosage, have dams more likely to die during pregnancy. Quantitatively, the invalid assumption by those performing the study is the assumption that maternal death is an independent variable to pup brain mass. 

It is also important to note that for the case where rare events might be the cause of a deviation from control values, a large enough sample must exist for sampling to average over the rare events. Since microcephaly may occur in only a small fraction of pups, the number of samples must be large compared to the inverse of the rate of its occurrence so that there are many such events in the experiment. For example, if microcephaly were to occur in only 1 per 100 pups, the number of samples must be large compared to 100 in order for the sampled average of Eq. \ref{Eq_2} to be reliably close to that of Eq. \ref{Eq_av}. Rare events that have large effect on brain mass, or have a distribution that is broad and therefore the statistical deviation is large, would not be correctly evaluated. 

For the rat toxicology experiment to be of use in understanding human toxicology, we must make the assumption that we can map the results of experiments on rats onto human beings in a way that is reliable. If we make this strong assumption, we would also take the rate of incidence of microcephaly in human beings as indicative of the rate in rats. The incidence of microcephaly in pregnancies in northeast Brazil in Pernambuco is approximately 30 per 10,000 births. We do not have information about the rate of exposure. Still, a $3\%$ rate would be consistent with a $31\%$ chance of having one case in 12 births. This suggests that only one of the dosage levels would be likely to have a single case of microcephaly. Discounting the $300$ mg/kg result is therefore inconsistent with the expectations based upon the incidence rate. The number of pups in each group is not sufficient for each of them to have affected individuals, if the effect occurs at the rate of microcephaly observed in Brazil.  A single individual could occur in any of the experimental groups, not necessarily in the high dosage one. (The uncertainty that the test results are applicable to human beings further reduces the reliability of the experiment as a test of toxicity.) 
 
We also note that in this study, the number of rat pups at the 1,000 mg/kg level is reduced to 78 from the target number 99. While the specific reason is not explained for this difference, there are two possible reasons in the study description. First, there are fatalities of the pregnant dams that led to adding an additional 12 dams to the study. Second, five dams were excluded from the study after the fact because of a mistaken feeding of pyriproxyfen starting on day 6 rather than day 7: ``Five animals (Nos. 2437 to 2441) in the 1000 mg/kg group were mistakenly administered the test substance from day 6 of gestation, and data of these animals were excluded from the evaluation."  It is unclear at what point they were excluded from the study. While each of these protocol changes might be appropriate, the choices that are made by experimenters that undermine statistical assumptions have become an increasing reason to question the reliability of studies \cite{reproduce0,reproduce1,reproduce2}.

Finally, in private communications Sumitomo also suggested that the use of relative brain weights may not be a good indicator of microcephaly compared to absolute brain weights. However, this statement is not reliably shown by prior studies \cite{Bailey2004,Tamaru1988}.




\subsubsection{Arhinencephaly} 

The case of Arhinencephaly in the rat experiment was suggested by a response of Sumitomo not to be indicative of microcephaly in humans. While both are malformations in the central nervous system, they may not be caused by the same mechanism. However, multiple types of neurological defects have been associated with the current epidemic of microcephaly in addition to microcephaly itself  \cite{Brasil2016}. While these have often been claimed to be associated with Zika infection, the existence of these alternative central nervous system defects can equally be associated with another cause, i.e. pyriproxyfen.

Sumitomo also affirms that the dose dependence argument applies in this case as well, as it is observed in the 300 mg/kg group but not in the 1000 mg/kg group in the rat teratogenicity study \cite{Ozaki1}. However, the existence of only one case of Arhinencephaly in the rat teratogenicity experiment must be interpreted as either a low probability event that is independent of the application of pyriproxyfen (random variable $X$ in Eq. \ref{Eq_2}) or a low probability event that is due to the application of pyriproxyfen (random variable $Y$). Sumitomo's argument assumes it is the former case rather than the latter case, without evidence for that claim. The assumption that it is not caused by pyriproxyfen is counter to the purpose of the testing which is to demonstrate that it does not cause harmful conditions. As stated before, the dose dependence argument does not apply to low probability events that occur at a rate that is low compared to the number of samples. 

\subsubsection{Rabbit study}

In a study conducted using rabbits, Sumitomo used a similar protocol to that for the rat study. Four groups were administered different amounts of pyriproxyfen, 0 mg/kg (control), 100mg/kg, 300 mg/kg, and 1,000 mg/kg. Rabbits in each group were mated, and resulting fetuses were examined for signs of toxicological damage. Examinations included skeletal variations, ossification of phalanges, and visceral anomalies \cite{Ozaki2}.  

The brain weight of fetuses was not measured in the rabbit teratology studies; instead, microcephaly was diagnosed by observation. When asked whether standards of observation have been sufficiently well established for the detection of microcephaly, Sumitomo responded that there is ``reported historical control data of the rabbit teratogenicity studies, so it suggests that microcephaly can be detected with the macroscopic observation even in rabbits" \cite{Ozaki2}.  

The response that it is possible to diagnose microcephaly in rabbits, does not answer the question as to whether the criterion in rabbits would be similar to the condition in humans, nor whether the number of cases being counted would correspond to the corresponding number in humans. This problem is particularly important where the incidence of microcephaly is rare, only a few percent even in the high incidence area of Brazil.  The criterion in human infants for microcephaly has been a key question in determining the number of cases that occur. The European Surveillance of Congenital Anomalies network (EUROCAT) uses more than three standard deviations below the normal head circumference. Brazil currently defines microcephaly as less than two standard deviations, although prior to November 2015, a broader definition was in use \cite{Yung2015}. Given difficulties defining microcephaly in humans, identifying microcephaly visually in a much smaller animal cranium may also be difficult. The identification of microcephaly is a key issue in evaluating whether observations in animal tests are useful. This is particularly relevant as the role of experimenter motivation has become increasingly recognized \cite{reproduce0,reproduce1,reproduce2}. While obtaining non-null statistically significant results is a motivation for scientific researchers, the opposite is true in a manufacturer conducted toxicology experiments. 

\section{Use of Pyriproxyfen in water supply in Brazil}

Pyriproxyfen has been used as an insecticide in northeast Brazil in response to an outbreak of dengue starting in the fourth quarter of 2014 \cite{Reis2016}. It is applied primarily to water storage containers used for home drinking water in areas that do not have a municipal water supply.

Pyriproxyfen has received widespread regulatory approval as an insecticide \cite{EPA,WHO2004}. Its primary use is on agricultural crops \cite{Devillers2013}, and against insects in households and on pets, e.g., on pet collars \cite{NRDC2000}. It is important to note that use of pyriproxyfen in drinking water on a large scale has never occurred anywhere else in the world, and was not used in Brazil prior to the fourth quarter of 2014. There have  been several small studies of the use of pyriproxyfen for mosquito control in villages. However, these studies did not monitor for toxicological effects. Prior tests were pilots for mosquito control in a few small communities in Malaysia \cite{Invest2008}, Peru \cite{Sihuincha2005}, Colombia \cite{Overgaard2012}, and Cambodia \cite{Seng2008}. These were small case studies, and only Cambodia and Colombia included a use in the drinking water supply \cite{Invest2008}. The research on impacts included measures of mosquito control but did not consider developmental effects in humans. Moreover, their scale was so small that birth defects occurring at a rate consistent with what is observed in Brazil would not have been observed. Human trials of pyriproxyfen in drinking water have not been carried out in a laboratory setting. Such experiments require a large participating sample in order to observe birth defects at such an incidence. Thus the conditions that are currently present in Brazil have not been replicated elsewhere, and the possibility that pyriproxyfen is the cause of the observed microcephaly cases cannot be ruled out by prior studies or experience with the use of pyriproxyfen.

Prior to 2014, Brazil primarily used temefos (an organophosphate) in water supplies to reduce mosquito populations. The switch to pyriproxyfen was made due to increasing temefos resistance in mosquitoes \cite{PdS2014}. Some areas of Brazil do not use pyriproxyfen, using Bacillus thuringensis (BT) toxin instead. 

It has been reported in the press that the Brazilian Ministry of Health \cite{MoHZika2016, MoHMicro2016} claims areas not using pyriproxyfen also report cases of microcephaly. For example, there is a press report that pyriproxyfen was not being used in Recife, and two other cities of the state of Pernambuco \cite{Romo2016, Bichell2016, NoAuthor2016}, though there are cases of microcephaly and Zika. This has been cited as a compelling reason for dismissing the possibility that pyriproxyfen is a cause of microcephaly \cite{NoAuthor2016}. According to the Dengue control office of Pernambuco \cite{personalCommunication}, pyriproxyfen is widely used in the Recife metropolitan area. An unconfirmed report \cite{CES1}, states that pyrixroyfen is not used in three specific municipalities (urban administrative areas)\textemdash Recife, Paulista, and Jaboatao do Guararapes, where Bt toxin is substituted. In this context it is important to recognize that pyriproxyfen exposure is relevant for the urban favelas and nearby rural areas where it would be used in water storage containers. It is relevant to urban areas with municipal water supply systems where such systems are augmented by water storage containers due to frequent failures \cite{Frontline}. Urban hospitals attract populations from a large area \cite{Douglas2016} and official reporting of cases is not broken down by municipality but rather by state, so that a direct analysis of geographical distribution of exposure to pyriproxyfen is difficult in these particular cases. A recent paper studied the association of microcephaly with the areas of claimed pyriproxyfen use and reports a lack of correlation \cite{Albuquerque16}. If the geographic distribution of use and the reported residences of mothers are confirmed, this study would provide strong evidence against pyriproxyfen as a cause. However, the study does not provide a reliable source for the geographic use of pyriproxyfen. Information about the birth location of favela residents (as opposed to more conventional urban area residents) is also not adequately confirmed. Both are necessary. Thus, further analysis is needed to determine whether the large number of cases in the state of Pernambuco have occurred in areas where pyriproxyfen is used. 

\section{Review of Zika incidence in Brazil and Colombia}

The existence of an epidemic of Zika in Brazil, the first observed in the western hemisphere, during the early part of 2015, preceding the observation of a large number of cases of microcephaly starting toward the end of 2015, led to the natural inference of a causal relationship between them. We review here existing evidence including geographic data about the incidence of Zika and microcephaly in Brazil and other countries \cite{Parens2016,NECSINov,r1,r2,August1,Bar-Yam2016}. A consistent picture suggests a low incidence of cases caused by Zika and another cause responsible for most of the microcephaly cases in Brazil. The possibility that a cofactor along with Zika infections is responsible is opposed by the low rate at which microcephaly cases are confirmed to be associated with Zika infections in Brazil. In particular, the ratio of microcephaly to Zika cases is inconsistent between government reported cases in Colombia and Brazil and among Brazilian states, where the majority of cases are confined to the northeast region. At the rate of microcephaly reported in Colombia, if all pregnancies in the Brazilian state of Pernambuco were infected by Zika, we estimate there would only be 105 cases of microcephaly in a year, whereas the number of confirmed cases is 386. The evidence against Zika as a cause implies other causes of microcephaly should be more fully considered, including pyriproxyfen. 

\subsection{Evidence linking Zika infection and microcephaly}

We begin by reviewing the evidence that is considered to link Zika with microcephaly, which we find is consistent with Zika being a cause of individual cases but does not associate it to the majority of microcephaly cases in Brazil. 

In May of 2016, the CDC declared Zika the cause of microcephaly in Brazil \cite{Rasmussen2016}. The CDC highlights the three factors that contributed to their conclusion:
\begin{itemize}
\item{Zika virus infection at times during prenatal development...consistent with the defects observed;} 
\item{a specific, rare phenotype involving microcephaly and associated brain anomalies in fetuses or infants with presumed or confirmed congenital Zika virus infection;} 
\item{data that strongly support biologic plausibility, including the identification of Zika virus in the brain tissue of affected fetuses and infants.}
\end{itemize}
The cited evidence is about individual cases and the primary evidence of a connection to the large number of cases observed is the timing of the initial Zika outbreak relative to the initial outbreak of microcephaly cases in Northeastern Brazil. 

A more extensive list of evidence linking Zika and microcephaly includes:
\begin{enumerate}
\item{Zika virus was detected in an aborted fetus with microcephaly after the mother had symptoms of infection in the 13th week of gestation. The virus was found in neurological tissue \cite{Mlakar2016}.} 
\item{Zika virus was observed infecting neural stem cells and affecting their growth \cite{Tang2016}.}
\item{Zika virus was found in the amniotic fluid of two pregnant Brazilian women, after both showed possible signs of infection, including fever and a rash \cite{Calvet2016}.}
\item{In a public release on February 12, 2016, Brazil's health ministry reported 462 confirmed cases of microcephaly or other alterations to the central nervous system, after investigation of 1,227 of 5,079 suspected cases of microcephaly recorded from October 22, 2015 until February 6, 2016. 3,852 remained under investigation. Brazil confirmed that 41 of these cases of microcephaly are combined with ``evidence of Zika infection...either in the baby or in the mother.'' It is unclear from this report what the Zika infection status is for the microcephaly cases for which evidence of Zika infection is not reported \cite{Nebehay2016,MdS2016}, i.e. whether they were investigated and had no evidence of infection or whether they were not investigated. The low percentage of confirmation $8.9\% = 41/462$ has increased to approximately $15.6\%$ \cite{bul7272016} in sporadic subsequent reports. } 
\item{In a release reported on November 14, 2015, the French Polynesian health authorities reported 17 (possibly 18) cases of central nervous system malformations, including 8-9 cases of microcephaly \cite{polynesia1,polynesia2,Cauchemez2016}. A retrospective study shows that a model in which first trimester exposure is a cause of microcephaly in French Polynesia is a good fit for the data \cite{Cauchemez2016}. The best fit model implies a $1\%$ incidence of microcephaly resulting from maternal exposure in the first trimester of pregnancy, providing strong statistical evidence for Zika as a cause, though the total number of cases is only 7 during the critical period. Including only live births would yield an even lower incidence: one microcephaly case over background.} 
\item{Zika is known to cause neurological damage in adults, typically leading to transitory paralysis, i.e. Guillian-Barre syndrome \cite{Cao-Lormeau2016,Vogel2016}.}
\item{A preliminary cohort report on outcomes of Zika infected pregnant women in Rio de Janeiro appears to provide strong evidence \cite{Brasil2016}. The study included 345 women, 134 Zika-infected pregnancies and 117 births. Multiple incidences of birth defects were determined either by ultrasound or at birth. Four cases of microcephaly were reported through ultrasonographic features. Two of these were associated with infections in the 5th and 6th months of pregnancy, inconsistent with both the French Polynesian analysis showing primarily early pregnancy effects and the absence of widespread cases in Colombia (below). Indeed, while 4 out of 23 pregnancies are reported as having anomalous ultrasounds when exposures happened in the first 4 months, 3 out of 9 (5 out of 9) pregnancies reported as infected in the 5th (6th or later) month have anomalies, including 2 still births in addition to the microcephalic ultrasounds \cite{error}. If similar exposure impacts were present in Colombia, a large number of anomalies, still births and microcephalic infants would have been reported very early during the epidemic there, and more by this time. The seemingly inconsistent two late term microcephaly cases were eventually reported as small for gestational age (SGA) rather than microcephaly at birth. While this might help resolve some of the discrepancy with other reports including Colombia, discounting the ultrasonographic findings leaves only one live microcephaly birth in the report. Finally, the Zika infected and small uninfected populations of the study are low/high income biased and therefore may also reflect environmental exposures of teratogenic agents associated with mosquito control that correlate with areas that have higher likelihood of infection. Perhaps most puzzling, the reported tracking of microcephaly through ultrasound during pregnancy in these two studies seems inconsistent with the absence of reports in other locations. Subsequent reports from Rio do not report nearly as high an incidence of microcephaly \cite{NECSINov}.} 
\item{A study of 32 microcephaly cases and 62 matched controls was carried out in 8 public hospitals of Recife, Pernambuco, Brazil between January and May 2016 \cite{Araujo16}. Immune factor and genetic (PCR) tests of Zika infection were performed. Of the microcephaly cases 13 (41\%) of 32 cases and none of 62 controls had laboratory-confirmed Zika virus infection. This provides strong confirmation that Zika causes microcephaly, but provides very limited information about the fraction of microcephaly cases in Brazil that are caused by Zika. If the negative test results are considered valid, the fraction of microcephaly that is not caused by Zika is substantial and the small size of the sample makes the fraction value uncertain. The statistical error range is consistent with the $15\%$ reported in government tests (point 4 above and Table \ref{Brazilratio}). Whether or not the negative results are conclusive is unclear, however this study shows that the best immunological and DNA tests available do not show that the large majority of microcephaly cases in northeast Brazil result from Zika infections.} 
\end{enumerate}

In summary, the central evidence that links Zika and microcephaly to individual cases is strong, but association to the large number of cases in Brazil is missing. 

\subsection{Analysis of geographic data on Zika and microcephaly}

\subsubsection{Number of microcephaly cases by country}

Zika infections have spread widely across Central and South America. Countries other than Brazil have reported a comparatively small number of cases ranging from a few to a few tens of cases \cite{WHO} (see Table \ref{sitrep}). Early in the time of the spread, it was unclear what delay should be present between the incidence of Zika and microcephaly related to the month of pregnancy of exposure. The number of actual Zika infections is also difficult to identify. However, since the spread of Zika occurred widely during the fall of 2015 and winter of 2016, the number of cases by the end of 2016 should be compatible with population numbers and rates of infection. While these are difficult to obtain, it is apparent from a cursory examination of the WHO reported  numbers of cases by country that there are no countries that rise to the level of the number of cases of microcephaly observed in Brazil, which has reached over $2,000$. Establishing the number of cases of Zika and the timing of those cases is needed to determine the relationship of Zika and microcephaly. We have performed a first analysis for the case of Colombia, the country that reports the second largest number of cases of microcephaly, as discussed in the following section.

\begin{table}[tb]
\begin{tabular}{|l|c|l|}
\hline
Country/Territory & Microcephaly Cases & Probable Location of Infection \\
\hline
Brazil & 2063 & Brazil \\ \hline
Cabo Verde & 9 & Cabo Verde \\ \hline
Canada & 1 & Undetermined \\ \hline
Costa Rica & 1 & Costa Rica \\ \hline
Colombia & 47 & Colombia \\ \hline
Dominican Republic & 10 & Dominican Republic \\ \hline
El Salvador & 4 & El Salvador \\ \hline
French Guiana & 10 & French Guiana \\ \hline
French Polynesia & 8 & French Polynesia \\ \hline
Grenada & 1 & Grenada \\ \hline
Guatemala & 15 & Guatemala \\ \hline
Haiti & 1 & Haiti \\ \hline
Honduras & 1 & Honduras \\ \hline
Marshall Islands & 1 & Marshall Islands \\ \hline
Martinique & 12 & Martinique \\ \hline
Panama & 5 & Panama \\ \hline
Paraguay & 2 & Paraguay \\ \hline
Puerto Rico & 2 & Puerto Rico \\ \hline
Slovenia & 1 & Brazil \\ \hline
Spain & 2 & Colombia, Venezuela \\ \hline
Suriname & 1 & Suriname \\ \hline
Thailand & 2 & Thailand \\ \hline
United States of America & 28 & Undetermined \\ \hline
\end{tabular}
\caption{Countries and territories that have reported microcephaly and/or central nervous system (CNS) malformation cases  \cite{sitrep,WHO}. While the WHO reports the microcephaly cases as associated with Zika infections, the association has not been shown in most cases.}\label{sitrep}
\end{table}

\subsubsection{Comparison of totals for Zika in Colombia and Brazil}

\begin{figure}[tb]
\centering
\includegraphics[width=17 cm]{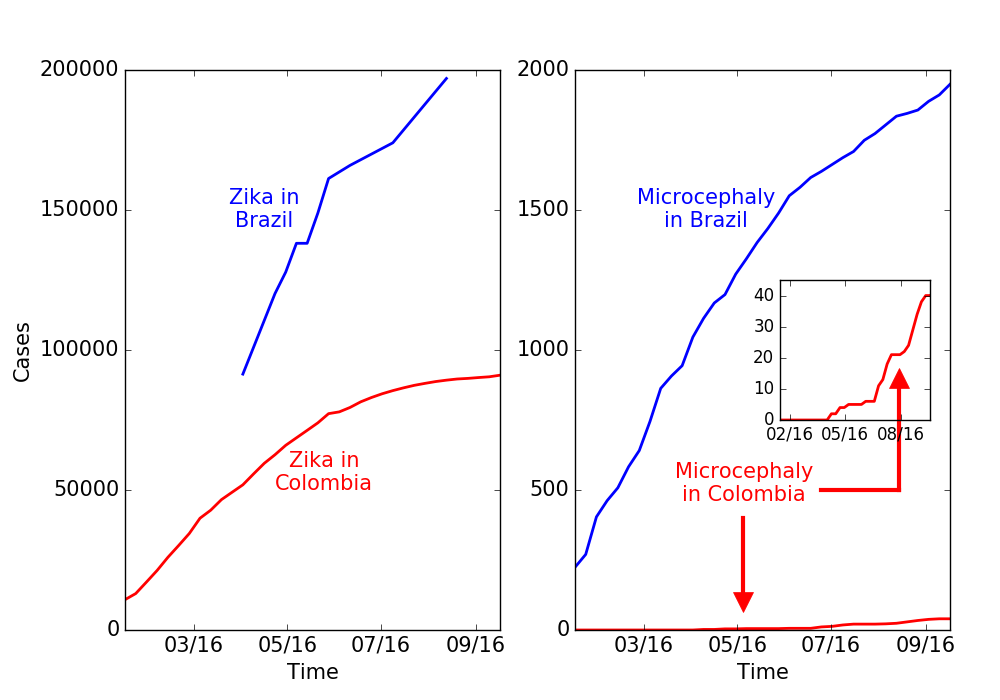}
\caption{Reported cases of microcephaly and of Zika in Brazil and Colombia. A. Cumulative reported cases of Zika in Brazil and Colombia. B. Total microcephaly cases reported in Brazil and Zika associated microcephaly cases reported in Colombia. The number of Zika cases  in Colombia is lower by a factor of 2, while the number of microcephaly cases is lower by a factor of 50. (Brazil reports total microcephaly numbers and does not distinguish those linked to Zika. Colombia reports only Zika-linked microcephaly cases. The historical background rate of microcephaly in Colombia is 140 per year.)}
\label{totals}
\end{figure}

Colombia reported a large number of Zika infections, but it has only seen a small number of Zika-associated microcephaly cases (see Fig. \ref{totals}). The number of Zika infections reported in Brazil is roughly 200,000 while that in Colombia is 90,000. The number of confirmed microcephaly cases in Brazil now exceeds 2,000, while the number reported by the government in Colombia linked to Zika is only 57. Moreover, despite many cases of Zika in other parts of Brazil, the majority of microcephaly cases have been confined to the northeast region, which has a population of approximately 50 million, comparable to that of Colombia. While questions remain about reliability of tests and reporting, the extent of the inconsistencies is difficult to account for. Overall, the discrepancies suggest other causes or co-factors, rather than Zika itself, are the primary source of microcephaly in Brazil. The timing of Zika and microcephaly cases in Colombia is discussed in the following section.

\subsubsection{Analysis of Zika caused microcephaly in Colombia}

\begin{figure}[tb]
\centering
\includegraphics[width=16 cm]{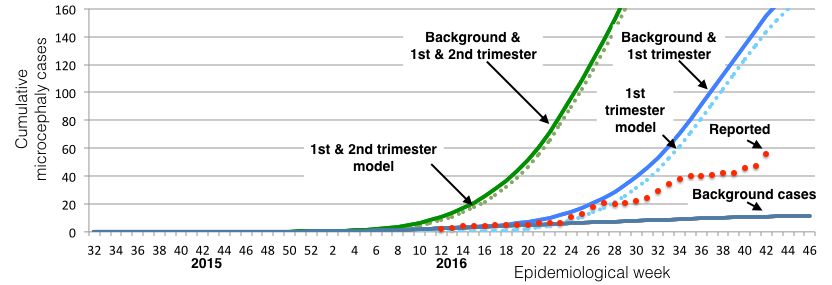}
\caption{Comparison of reported cases in Colombia with background and Zika causal models. Reported cases of Zika and microcephaly (red dots) are compared with expected number of background cases due to coincidence of microcephaly with Zika infections at a rate of 2 per 10,000 births (gray) and two models of Zika-induced microcephaly suggested by the study in French Polynesia \cite{Cauchemez2016}. The first assumes that pregnancies infected in the first trimester have a 1:100 rate of microcephaly; the second assumes that pregnancies infected in the first and second trimester have a 1:200 rate of microcephaly. The data is consistent with just background cases until the report of epidemiological week 24, ending June 18. The data in weeks 24 through 27 is reasonably consistent with the first trimester model, but not with later weeks. The overall rate is much lower than that consistent with the number of cases in northeast Brazil.}
\label{fig:Colombia}
\end{figure}

We construct a model of the time period of infection during pregnancy that results in microcephaly in Colombia based upon available data on Zika and microcephaly (Fig. \ref{fig:Colombia}).  We find that of the 56 confirmed microcephaly cases reported as of epidemiological week 42 \cite{bul}, only about 45 cases can be attributed to Zika infections while 11 are due to the background effect of random cases of microcephaly that by coincidence occur in Zika infected pregnancies. Colombia is not reporting the number of microcephaly cases that are unlinked to Zika. 
 
The Colombian outbreak of Zika began in August of 2015 and the number of infections increased rapidly in early 2016. The outbreak diminished and was declared over by July, 2016 \cite{ColombErad}. Many of the exposed pregnancies should result in births in the second and third quarters of 2016. We performed a quantitative analysis based upon an extensive preliminary cohort study published on June 15, 2016 in the New England Journal of Medicine \cite{ColombiaNEJM}. The study reports results of women infected until March 28, 2016, whose pregnancies were observed until May 2, 2016. The study tracked 1,850 women, whose date of infection with Zika is known relative to the start of the pregnancy. Of these, 532, 702 and 616 were infected in the first, second and third trimesters respectively. 16\%, 29\% and 93\% (85, 204 and 583) of the pregnancies concluded. No cases of microcephaly were observed. The total number of pregnancies with Zika infections is much larger, with 11,944 cases with Zika symptoms being observed in clinical settings. At the time of the article, no cases of microcephaly occurred in any of nearly 12,000 pregnancies. However, the report cites 4 cases of microcephaly with Zika in the general population that did not report any Zika symptoms, these cases being reported prior to April 28. This finding implies that there are many more unreported, asymptomatic, cases of Zika infection. Since there is less than a 1 in 12,000 incidence of Zika until this point in the epidemic, there should be at least 4 times as many infected individuals that do not have symptoms in order for there to be 4 microcephaly with Zika cases---a ratio of total and symptomatic cases of 5:1---for a total of 60,000 Zika cases ($5 \times 12,000$). There is other evidence for underreporting of Zika. For example, the incidence among women is twice that reported in men. Women at home may be infected at a higher rate, or reporting is focused on women because of the concern about Zika's maternal effects.

It is useful to have a reference model of Zika as a cause of microcephaly even though there is no consistency between different observational studies. For this purpose, we adopt as a reference a model that comes from the study reported in French Polynesia \cite{Cauchemez2016}, which provided evidence that 1 in 100 pregnancies exposed in the first trimester, or, alternatively, 0.5 in 100 of all pregnancies exposed in the first and second trimester, resulted in microcephaly. This study was based upon a small number of cases and ultrasound detection rather than births. Seven cases of microcephaly were reported above background, but only one of them was a birth; the others were detected by ultrasound. The high rate of ultrasound detection is not consistent with observations by ultrasound of microcephaly in other countries, including Colombia. We use the model for French Polynesia as a reference model for comparison with other data in order to clarify which data is or is not consistent with that model. We note that this rate is about 100 times larger than the minimum reported background rate for microcephaly, 2 in 10,000 \cite{cdc1} if all pregnancies are infected. The Zika-induced cases should be considered to be in addition to the background cases, which are due to other causes. It should be emphasized that the French Polynesia model is not necessarily compatible with the observed cases in Brazil; and has not been adequately validated in any context including French Polynesia.

We construct a model of the Zika-infected pregnancies by considering each pregnancy to have a uniform probability of infection across 39 weeks. This enables us to estimate the total number of Zika-infected pregnancy births as well as the number that are born after exposure in the first trimester or in the first and second trimesters. The total number of cases should be a combination of those with Zika exposure and background cases. For background cases, any birth has a probability of 2 in 10,000 of microcephaly. If a Zika infection occurred anytime during pregnancy, it would be a Zika and microcephaly case at birth.

Fig. \ref{totals} shows reported cases of microcephaly linked to Zika infections (red dots). These are compared with background cases predicted based on the number of Zika-infected pregnancies \cite{ColombiaNEJM} and two models of Zika as a cause of microcephaly that originate from the outbreak in French Polynesia.

The cases of Zika and microcephaly reported until June 11, 2016 are consistent with the expected background rate of birth defects that would have occurred in those infected with Zika, even if Zika were not a cause \cite{r1}. After that date, cases initially tracked a trajectory matching a model of 1\% of pregnancies infected in the first trimester developing microcephaly \cite{r2}. However, the number of cases subsequently plateaued and increased in stages but more slowly than expected from the model. As of epidemiological week 42, a total of 56 cases were reported \cite{bul}, 11 of which can be explained as background cases unrelated to Zika. This is much less than the 155 total cases predicted from 1\% of first trimester pregnancies along with the background rate. It is also far less than the over $2,000$ cases reported in Brazil. Based on this data, approximately $0.23\%$ of first trimester infections ($0.075\%$ of infected pregnancies) in Colombia have resulted in microcephaly. The coincidence of increased microcephaly cases with the timing of births infected in the first trimester suggests Zika is responsible for a limited number of microcephaly cases. This is consistent with other reports that link Zika with a few cases of microcephaly, as has also been shown with other viral infections, but is not consistent with Zika being the cause of the majority of cases reported in northeast Brazil. 

\begin{figure}[tb]
\centering
\includegraphics[width=15 cm]{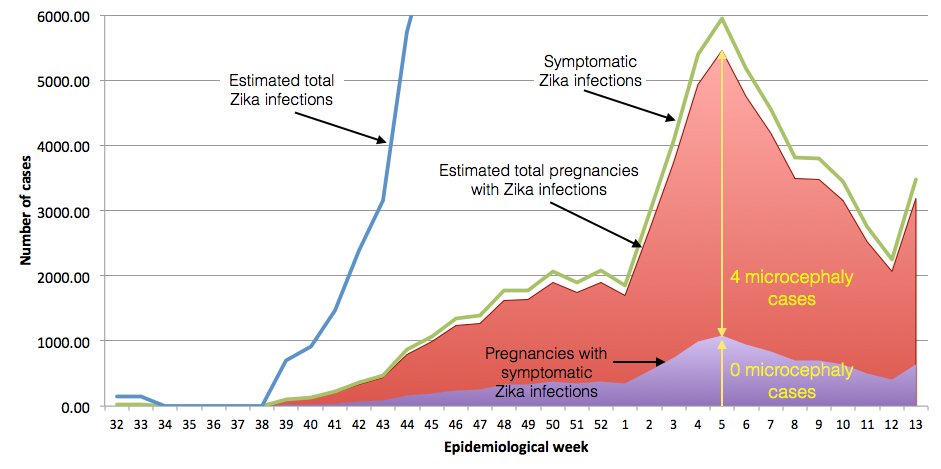}
\caption{Population of Zika infected pregnancies. To model the Colombian Zika and microcephaly epidemic, the reported number of symptomatic cases per week until March 28 (green line) \cite{ColombiaNEJM} is normalized by the number of reported Zika infected pregnancies (11,944, shaded purple) and multiplied by 5 to obtain the total number of symptomatic and asymptomatic cases (red shading), due to the observation of four Zika and microcephaly cases that do not have Zika symptoms. While we don't use it directly, the total number of Zika infections can be estimated (blue line) by similarly multiplying the number of reported cases, correcting for the bias in reporting between women and men, assuming the infection rate is comparable. Other assumptions about the total number of cases do not affect the results reported here.}
\label{Pop}
\end{figure}

\begin{figure}[tb]
\centering
\includegraphics[scale=0.38]{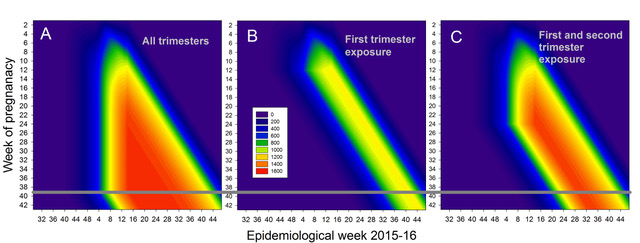}
\caption{Pregnancies during a particular week (vertical axis) present during a particular epidemiological week (horizontal axis). (A) Total number of pregnancies, (B) the number of pregnancies exposed in the first trimester, and (C) the number of pregnancies exposed in either the first or second trimester. A uniform exposure by week of pregnancy is assumed. The number of births in each category is approximately the number that crosses the 39/40 week boundary (horizontal gray line), but it is more accurately given by weighting them according to the distribution of births from week 35 through 43.}
\label{Pregnancies}
\end{figure}

Details about the construction of the model for Zika-caused microcephaly are provided in Figs. \ref{Pop} and \ref{Pregnancies}. A similar model has previously been applied to a single municipality of Bahia \cite{Physicians2016}, without discussion of comparisons with other areas. Other simulations of Zika as a cause of microcephaly also do not achieve numbers consistent with those reported \cite{Zhang2016}.

\subsubsection{Brazil microcephaly and Zika geographic data}

In addition to the differences between Colombia and Brazil, questions have been raised about the geographic distribution of microcephaly cases across Brazil. Much of the original data on the Zika infection did not include the numbers of individuals infected and only identified whether or not a state was infected by Zika. The mild symptoms of the infections in many individuals make precise counting difficult. Reporting was not required of states by the national government until 2016. However, since the beginning of 2016 better reporting is available, though the reliability of the numbers can be challenged. Nevertheless, informal reports suggest that even without precise numbers the existing data are difficult to reconcile with expectations based upon Zika as a cause of microcephaly: The spread of Zika across different states of Brazil has not been accompanied by a comparable wave of microcephaly \cite{naturebutler, McNeil2016, Phillips2016}.

\begin{figure}[tb]
\centering
\includegraphics[width=16 cm]{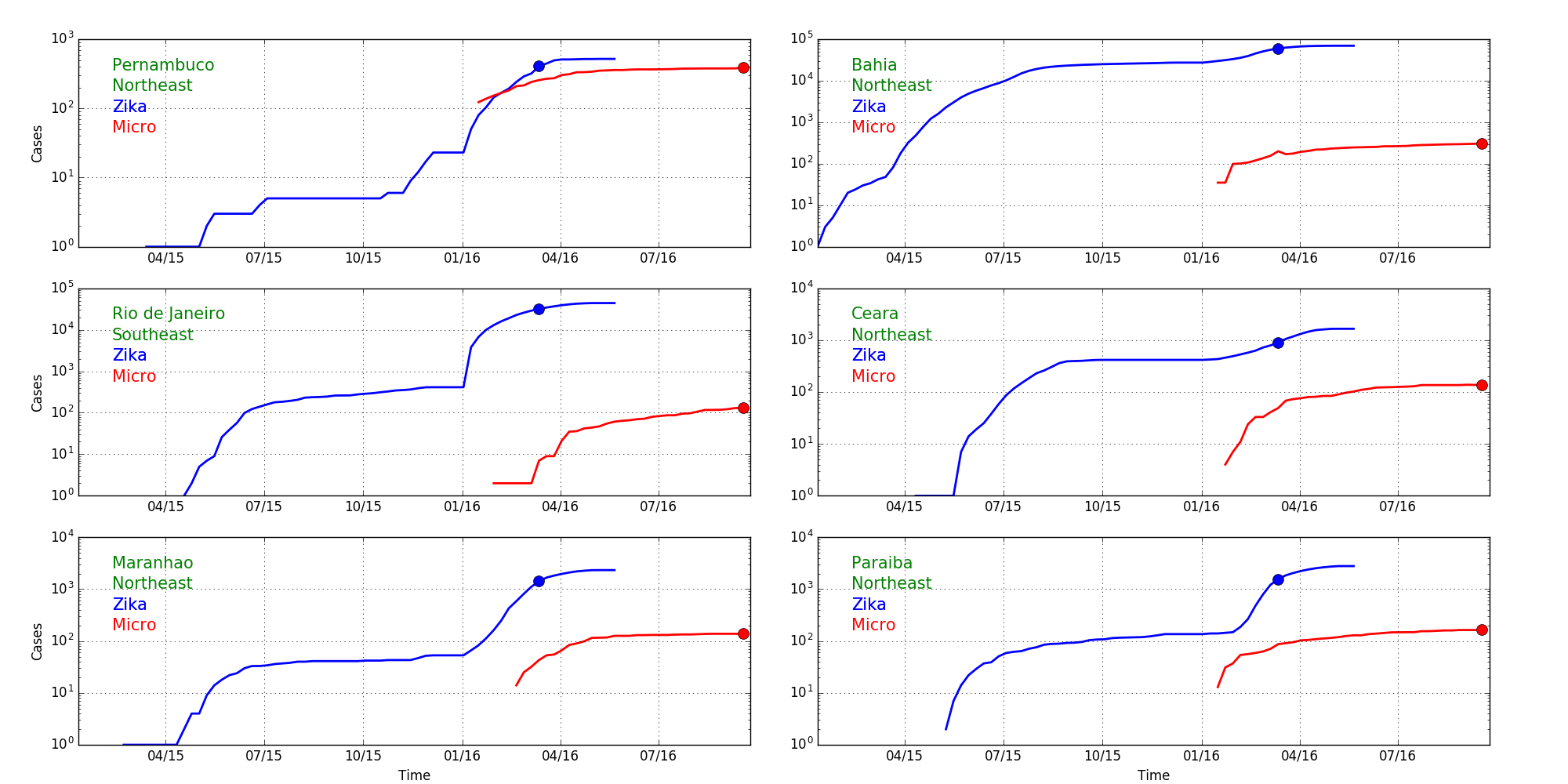}
\caption{Cumulative Zika (blue) and microcephaly (red) cases over time in five northeast Brazilian states and the state with the largest number of cases elsewhere, Rio de Janeiro (log scale). Blue and red dots are separated by 33 weeks, the expected delay between first trimester infections and expected microcephaly births caused by them. The differences in ratios in different states (Fig. \ref{ratio}) suggests that Zika is not the cause of microcephaly.}
\label{logscale}
\end{figure}

Quantitative comparisons are made easier by the analysis of Colombia which shows that the cases of Zika-caused microcephaly occur only due to infections in the first trimester, narrowing the window of delays between exposure and case reports of microcephaly. We can therefore compare reports of Zika infections with microcephaly reports 33 weeks later to identify the potential causal relationship between Zika and microcephaly in different states of Brazil. Zika and microcephaly cases reported in six Brazilian states in 2016 are shown in Fig. \ref{logscale} along with an indicator of the time difference of 33 weeks. 

We see that the ratio between Zika and microcephaly reports varies between $1$ and approximately $1/1000$. That it is possible to have a ratio of about $1$ is surprising if one views Zika as a cause of microcephaly. The number of pregnancies with Zika infections should be much less than the number of Zika cases. To estimate the number of Zika infected pregnancies in the first trimester at a particular time we would multiply the number of reported cases by an underreporting factor of 5 \cite{r1} to obtain an estimate of the actual number of cases, and multiply by the fraction of the population that is pregnant in the first trimester at any time, $0.37\%$ (birthrate per day times 90), so the number of susceptible pregnancy Zika infections would be about $1.8\%$ of the number of Zika cases. This number is smaller than the number of microcephaly cases in several states. 

As a preliminary upper bound on the number of Zika induced microcephaly cases we might consider the incidence from Colombia of $0.075\%$ of pregnancies and calculate the number of cases that would be present if everyone was infected by Zika. For Pernambuco, with a population of $9.3$ million and an approximate birth rate of $15$ per $1,000$, the number of microcephaly cases would be approximately $105$ a year, much less than the actual number $386$.

Moreover, the inconsistency among the states is independent of any calculation of the rates of the number of susceptible pregnancies. The wide range of values suggests that Zika is not the primary cause of microcephaly. All calculations are sensitive to the possibility that reporting is poor and inconsistent across states and countries. However, it remains difficult to identify a way to reconcile the extent of the inconsistency across Brazil. 

\begin{figure}[tbp]
\centering
\includegraphics[width=14 cm]{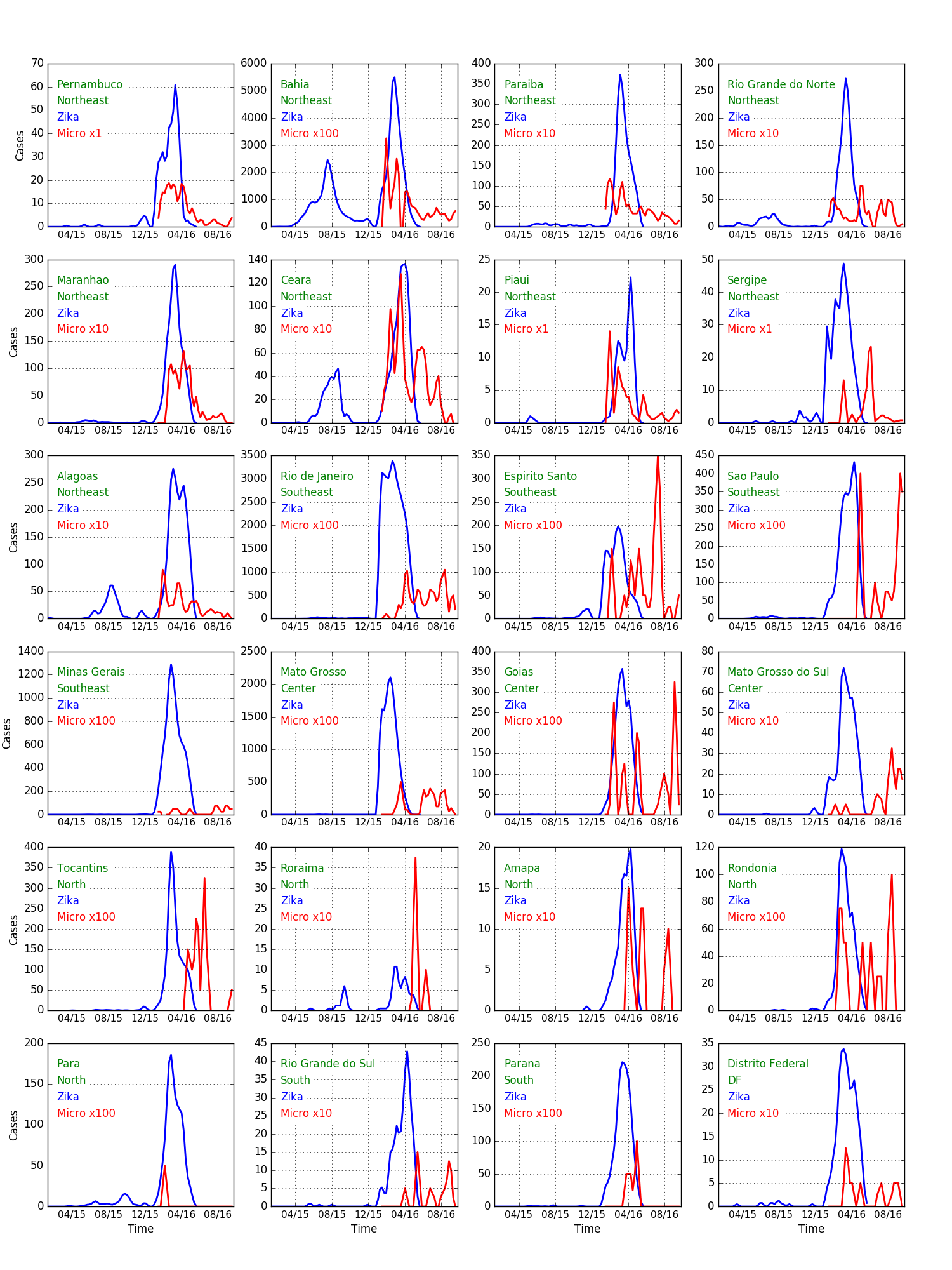}
\caption{Weekly Zika (blue) and microcephaly (red) reports over time in each Brazilian state. Note the multipliers for the microcephaly numbers. If Zika is the cause of the cases of microcephaly a delay of about 33 weeks should be seen between peaks of the former and latter. This appears to be the case for Bahia, Ceara, and Alagoas for early peaks of Zika and later peaks of microcephaly. We note that if seasonal use of insecticides coincides with outbreaks, then the cause may also be those insecticides. A filter \{0.25, 0.5, 0.25\} has been applied to smooth the data.}
\label{peaks}
\end{figure}

\begin{figure}[tb]
\centering
\includegraphics[width=12 cm]{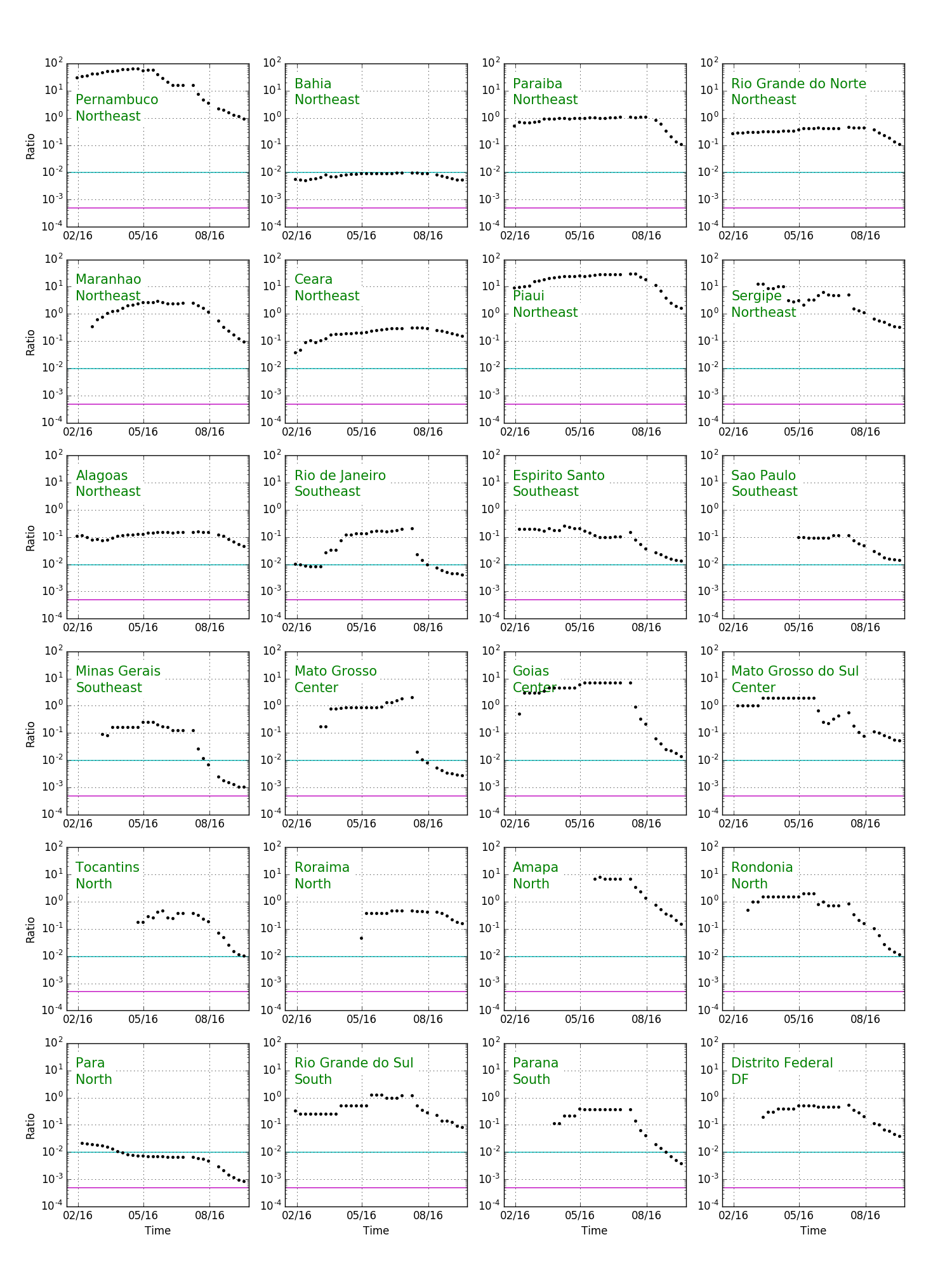}
\caption{Ratio of microcephaly cases to Zika cases 33 weeks earlier as a function of time for all Brazilian states. The horizontal green line represents a ratio of $1\%$ and the purple horizontal line represents a ratio of $0.05\%$. To obtain the number of microcephaly cases per first trimester Zika infected pregnancy (rather than all Zika cases) we would have to include both unreported Zika cases and multiply by the proportion of pregnancies, multiplying the ratio by a factor of 25. Differences between rates would remain.}
\label{ratio}
\end{figure}

Expanding the discussion to all states of Brazil, Fig. \ref{peaks} shows confirmed cases of Zika (blue) and microcephaly (red) for Brazilian states. The widely-varying relative proportion of Zika and microcephaly is apparent in the multipliers used to show the microcephaly data on the same vertical scale. Fig. \ref{ratio} shows the ratio between Zika and microcephaly cases as a function of time (including the 33 week delay) for all Brazilian states. As the figure shows, ratios vary widely across the country, but are higher than the small proportion of government reported cases in Colombia. 

Table \ref{ratiomicrozika} shows the ratio of microcephaly cases to Zika cases reported 33 weeks previously for all Brazilian states, including cases as of October, 2016, and the maximum over the year. In the northeastern states, only Bahia is reporting a number of cases consistent with Zika as a primary cause of microcephaly. Interestingly Rio de Janeiro, the only state outside the northeast that has more than 100 cases, is also consistent with Bahia (though the ratio was higher at earlier times). Other states outside the northeast have too few cases to reliably compare.

We note that a previously-published report on the total number of Zika and microcephaly cases across all states, in effect, inappropriately linked the cases of Zika reported for Bahia with the cases of microcephaly reported for Pernambuco \cite{Faria16}. 

\begin{table}[tb]
\begin{footnotesize}
\begin{tabularx}{\textwidth}{|X|X|X|X|X|X|X|X|}
\hline
Region & State & Current Zika & Current Microcephaly & Current Ratio & Zika at Maximum Ratio & Microcephaly at Maximum Ratio & Maximum Ratio \\
\hline
Northeast & Pernambuco & 408 & 386 & 0.95 & 5 & 334 & 67 \\ \hline
Northeast & Bahia & 59560 & 312 & 0.0052 & 27290 & 268 & 0.0098 \\ \hline
Northeast & Paraiba & 1547 & 166 & 0.11 & 140 & 155 & 1.1 \\ \hline
Northeast & Rio Grande do Norte & 1295 & 138 & 0.11 & 264 & 123 & 0.47 \\ \hline
Northeast & Maranhao & 1419 & 138 & 0.097 & 43 & 126 & 2.9 \\ \hline
Northeast & Ceara & 906 & 137 & 0.15 & 431 & 136 & 0.32 \\ \hline
Northeast & Piaui & 59 & 99 & 1.7 & 3 & 89 & 30 \\ \hline
Northeast & Sergipe & 384 & 123 & 0.32 & 2 & 26 & 13 \\ \hline
Northeast & Alagoas & 1766 & 84 & 0.048 & 504 & 79 & 0.16 \\ \hline
Southeast & Rio de Janeiro & 31542 & 130 & 0.0041 & 412 & 87 & 0.21 \\ \hline
Southeast & Espirito Santo & 1727 & 23 & 0.013 & 23 & 6 & 0.26 \\ \hline
Southeast & Sao Paulo & 1779 & 26 & 0.015 & 88 & 10 & 0.11 \\ \hline
Southeast & Minas Gerais & 7539 & 8 & 0.0011 & 12 & 3 & 0.25 \\ \hline
Center & Mato Grosso & 16680 & 47 & 0.0028 & 17 & 35 & 2.1 \\ \hline
Center & Goias & 1721 & 24 & 0.014 & 2 & 14 & 7.0 \\ \hline
Center & Mato Grosso do Sul & 370 & 20 & 0.054 & 1 & 2 & 2.0 \\ \hline
North & Tocantins & 1712 & 18 & 0.011 & 23 & 11 & 0.48 \\ \hline
North & Roraima & 62 & 10 & 0.16 & 21 & 10 & 0.48 \\ \hline
North & Amapa & 59 & 9 & 0.15 & 1 & 8 & 8 \\ \hline
North & Rondonia & 599 & 7 & 0.012 & 2 & 4 & 2.0 \\ \hline
North & Para & 1139 & 1 & 0.00088 & 47 & 1 & 0.021 \\ \hline
South & Rio Grande do Sul & 123 & 10 & 0.081 & 4 & 5 & 1.3 \\ \hline
South & Parana & 1014 & 4 & 0.0039 & 10 & 4 & 0.40 \\ \hline
DF & Distrito Federal & 201 & 8 & 0.040 & 11 & 6 & 0.55 \\
\hline
\end{tabularx}
\caption{Ratio of microcephaly cases to Zika cases reported 33 weeks previously for all Brazilian states, including cases as of October, 2016, and the maximum over the year.\label{ratiomicrozika}}
\end{footnotesize}
\end{table}

\subsubsection{Analysis of Bahia, Brazil}

\begin{figure}[tb]
\centering
\includegraphics[width=14 cm]{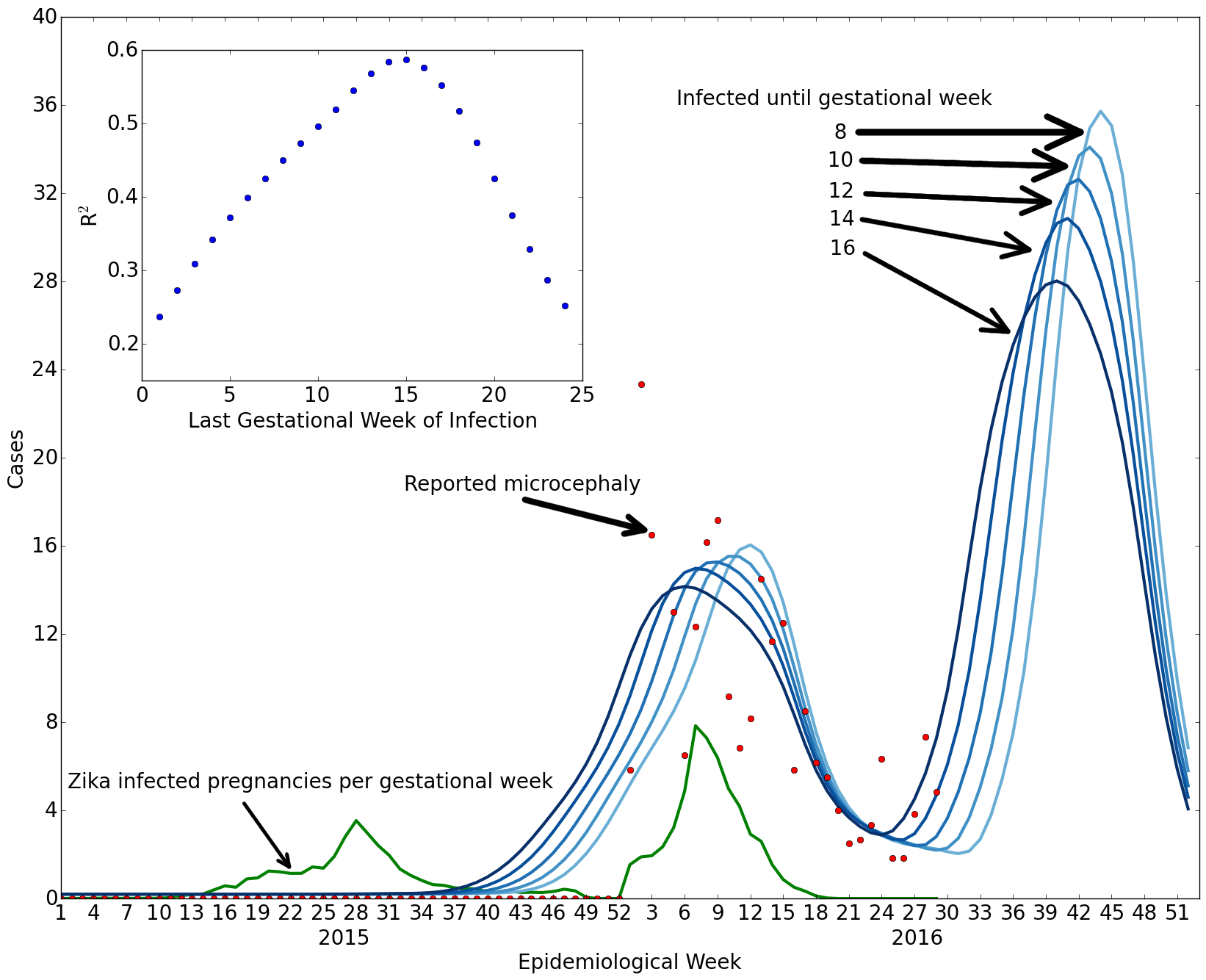}
\caption{Comparison of Zika and microcephaly cases in Bahia, Brazil. We compare reported microcephaly cases with or without symptoms of Zika (red dots) with the projected number of cases for susceptibility up to weeks 8, 10, 12, 14, 16 (shades of darker blue) with microcephaly in 84\%, 71\%, 63\%, 56\%, 49\% of the Zika pregnancies (optimized fit), based upon the reported number of Zika infections (green shows estimated Zika infected pregnancies per week of gestation). The left green peak  gives rise to the central red/blue peak due to the delay between first trimester infection and the births. The inset shows the goodness of fit for different numbers of weeks in which infections cause microcephaly. The best fit overall occurs for susceptibility up to 16 weeks, but the initial rise is more consistent with 8 weeks, suggesting that the Zika-induced microcephaly cases are likely due to infections approximately during the second month of pregnancy rather than the first trimester as a whole.}
\label{micro}
\end{figure}

Compared to other states in northeast Brazil, Bahia reports the lowest proportion of microcephaly cases relative to Zika cases and has the second highest number of microcephaly cases. We might speculate that the reported number of Zika cases may be more reliable and a higher proportion of cases may be caused by Zika. In August 2016, we analyzed the data available for the state of Bahia [Fig. \ref{micro}]. We observed that Bahia has a peak of microcephaly whose timing relative to a peak in Zika appears to be consistent with Zika as a cause. Nevertheless, our analysis suggests that the incidence of microcephaly is much higher than in either French Polynesia or Colombia. We compare three numbers: the number of confirmed cases, 263, the number of suspected cases, 1154, and the number confirmed as having both Zika and microcephaly, 41. The first two are reported from Bahia, while the last is taken from the national ratio of confirmed Zika and microcephaly cases as a fraction of confirmed microcephaly cases, $15.6\%$.

\begin{figure}[tb]
\centering
\includegraphics[width=16 cm]{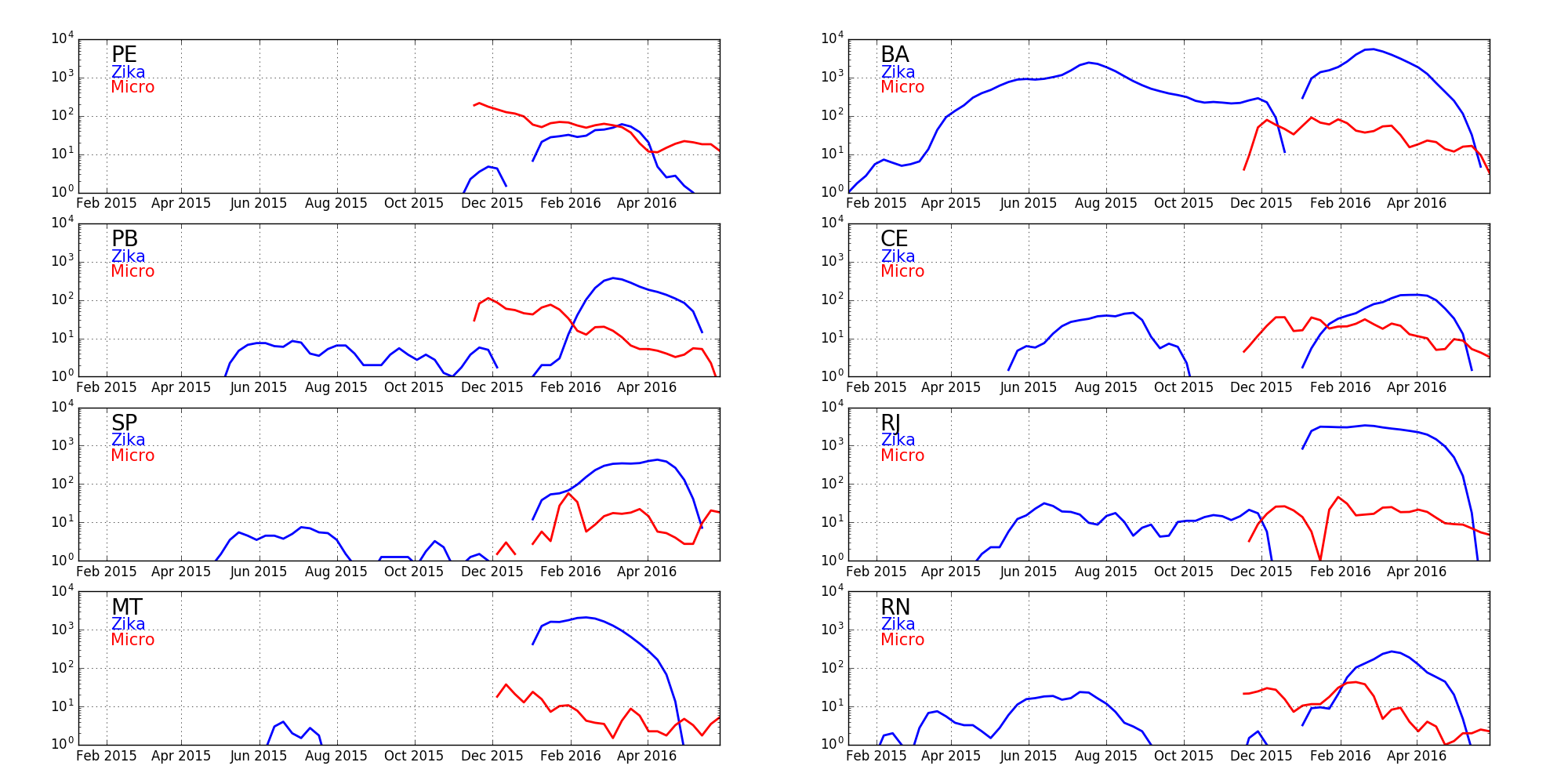}
\caption{Plots of Zika and microcephaly cases in Brazil. Plots (log scale) are shown for eight states of Brazil: Pernambuco (PE), Bahia (BA), Paraiba (PB), Cear\'a (CE), S\~ao Paulo (SP), Rio de Janeiro (RJ), Mato Grosso (MT), and Rio Grande do Norte (RN).  Bahia has the highest counts of Zika relative to microcephaly six months later when births of pregnancies exposed in the first trimester are expected.}
\label{states}
\end{figure}

\begin{figure}[tb]
\centering
\includegraphics[width=16 cm]{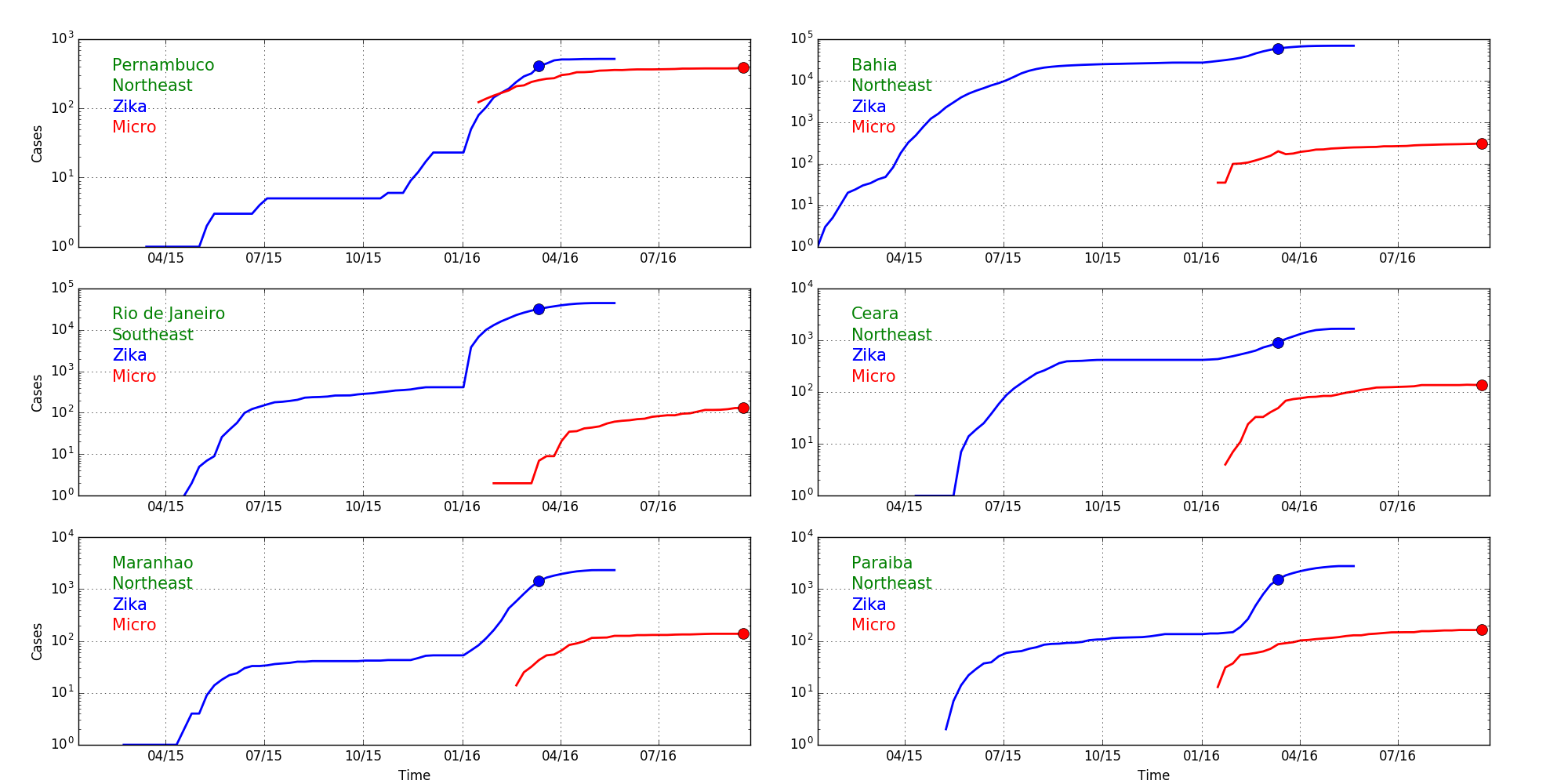}
\caption{Same as Fig. \ref{states}, but using a linear scale.}
\label{states1}
\end{figure}

According to the data that are available, the fraction of first trimester pregnancies exposed to Zika that have confirmed microcephaly is $63\%$, the number of suspected cases is $289\%$ of first trimester Zika-exposed pregnancies (which in principle is not inconsistent 
with a first trimester model of confirmed microcephaly cases), and the fraction of pregnancies exposed to Zika that are confirmed to have both Zika and microcephaly is $9.8\%$. All of these values are significantly larger than the $1\%$ rate obtained from French Polynesia and used above to model Colombia. We note that the population of Bahia is about 10 million, or 1/5 of the total population of Colombia. The relative reliability of Zika case reporting is unclear. In order for the microcephaly fraction to correspond to the $1\%$ first trimester model of French Polynesia, the number of Zika cases would have to be underreported in Bahia by a factor of 63. A large discrepancy in the Zika reporting rates would be necessary if Zika is consistently causing a certain percentage of microcephaly. Our results are dominated by the analysis of Zika reporting in 2015, yet reporting only became mandatory in 2016. In any case, the analysis shown in Figs. \ref{states} and \ref{states1} suggests that the second peak of the Zika epidemic would be expected to give rise to a new set of microcephaly cases. 

The seemingly large discrepancy between microcephaly counts in Colombia and Bahia echoes discrepancies between different parts of Brazil that led to questions about whether there are additional factors that affect the microcephaly rates \cite{naturebutler}.

\begin{table}
\begin{tabularx}{\textwidth}{|X|X|X|X|}
\hline
State & Confirmed Zika \& microcephaly & Confirmed microcephaly & Suspected microcephaly \\
\hline
Bahia (BA)	& 9.8\%	& 63\%	& 289\%	\\
Rio de Janeiro (RJ)	&	318\%	&	\textgreater1000\%	&	\textgreater1000\%\\
Cear\'a	(CE) &	349\%	&	\textgreater1000\%	&	\textgreater1000\%\\
Rio Grande do Norte (RN)	& 644\%	&	\textgreater1000\%	&	\textgreater1000\%	\\
Para\'iba (PB)	&	\textgreater1000\%	&	\textgreater1000\%	&	\textgreater1000\%	\\
Pernambuco (PE)	&	\textgreater1000\%	&	\textgreater1000\%	&	\textgreater1000\%	\\
S\~ao Paulo	(SP) &	\textgreater1000\%	&	\textgreater1000\%	&	\textgreater1000\%	\\
Mato Grosso (MT)	&	\textgreater1000\%	&	\textgreater1000\%	&	\textgreater1000\%	\\
\hline
\end{tabularx}
\caption{Rate of microcephaly in pregnant women infected with Zika during the first trimester for the eight Brazilian states with the largest numbers of microcephaly cases.}
 \label{Bahiatable}
\end{table}

\subsubsection{Ultrasound versus birth identification}

As part of the analysis of cases in Colombia and Brazil and other locations, it is worth noting that there has been an inconsistency among medical reports about the rate at which ultrasound detects Zika induced birth defects including microcephaly. There may be many reasons for such inconsistencies but an easy resolution of them is lacking. The French Polynesia study was based primarily upon cases detected by ultrasound. This is inconsistent with the reports from Colombia where only one case has been officially reported \cite{ColombiaNEJM}. A preliminary cohort study report in Rio \cite{Brasil2016} reported a $29\%$ rate of fetal abnormalities among Zika infected pregnancies including $10\%$ microcephaly cases based upon ultrasound findings of all exposed pregnancies including those in the third trimester. It is hard to reconcile this study with either the French Polynesia study, with only approximately $1\%$ microcephaly for first trimester exposures, or reports from Colombia. Indeed it is hard to understand how this study can be made consistent with other general findings given the rate of birth defects, still births and microcephaly. Most essentially, it is unclear why ultrasound-based observations of microcephaly in Colombia have not been more widely-reported given the 12,000 confirmed Zika cases there as of March 28, 2016. 

\subsubsection{Confirmed cases of Zika and microcephaly in Brazil}

One of the central questions about the role of Zika in Brazil is the absence of confirmation of cases of microcephaly as having Zika infections in government reporting (see Table \ref{Brazilratio}). The tests being performed are not sufficiently specified, and their reliability is unknown. Reports for Brazil have not been provided in recent months. While DNA tests performed sufficiently long after an infection might not be positive due to clearing of virus, it is unclear why immunological tests would not yield higher rates of confirmation. The immunological tests are likely to have false positives due an inability to fully distinguish Zika from Dengue and, perhaps, Chikungunya infections. On the other hand if it is assumed that Zika is the cause, then the $15.6\%$ confirmation rate implies there are many false negative results, the reason for which are not discussed. Absence of evidence of Zika infection may therefore be an indication that Zika is not the cause. What has not been observed are high rates of confirmation of Zika infection that would yield strong confidence that Zika is the cause of most cases. 

The absence of confirmation of Zika infections in Brazil is particularly challenging given the results of the French Polynesia study \cite{Cauchemez2016}. In that study, the population had seropositive results in $66\pm4\%$ ($95\%$ confidence interval) at the end of the epidemic. Thus there is very limited possibility of undetected cases there, suggesting the serological tests have high enough reliability to consider the negative results in Brazil as indicating that those cases are not the result of Zika infections. (We note that to the limited extent that the actual Zika infection rate is higher in French Polynesia, the incidence rate of microcephaly per maternal infection would have to be reduced.)   

\begin{table}[t]
\begin{tabularx}{0.8 \textwidth}{|X|X|X|X|}
\hline
Date & Confirmed Zika and microcephaly & Total microcephaly & Ratio \\
\hline
February 12, 2016 & 41 & 462 & 8.9\% \\ \hline
March 1, 2016 & 82 & 641 & 12.8\% \\ \hline
March 29, 2016 & 130 & 944 & 13.8\% \\ \hline
July 20, 2016 & 267 & 1709 & 15.6\% \\ \hline
July 27, 2016 & 272 & 1749 & 15.6\% \\ \hline
\end{tabularx}
\caption{The dates and numbers of confirmed Zika cases within the set of total cases of microcephaly in Brazil \cite{MdS2016,bul312016,bul3292016,bul7202016,bul7272016}.\label{Brazilratio}}
\end{table}

\subsubsection{Additional unconfirmed microcephaly cases in Colombia}

When considering the implication of the Brazilian microcephaly count for the Colombian ones, we note that the models in Fig. \ref{micro} consider all Zika-induced microcephaly cases as being confirmed to have Zika infections. Ignoring the serological tests in French Polynesia, we might consider speculatively that there are many more microcephaly cases whose Zika infections are undetected as they were not in Brazil. If the confirmation rate is similar to the $15.6\%$ found in Brazil, we would multiply the number of confirmed Zika induced microcephaly cases by a factor of $6.4 = 1/15.6\%$ to obtain the total number of Zika induced microcephaly cases. We would then have to increase the rate of microcephaly by this factor, which would make the Colombian results have a total of approximately $6.4*45=288$ cases, which would be a higher rate than the French Polynesian results (raising the question as to why many more cases were not observed in French Polynesia), but somewhat more consistent with the much higher microcephaly rates in Bahia. The additional cases (of order 300) should also be observed in microcephaly cases above background, but without evidence of Zika infection. 

Alternatively, if we use the estimated rate of microcephaly induced by Zika from French Polynesia of 1 in 100 pregnancies exposed in the first trimester \cite{Cauchemez2016}, there should be 200 microcephaly cases arising from Zika exposure of pregnancies infected until March 28. However, only about 1 in 6.4 of these, or 30, would also have confirmed Zika infections, if this fraction follows the pattern in Brazil. Reports of confirmed microcephaly cases not linked to Zika are not being published by the Colombian authorities, though we might consider the possibility that they have been included in the cases that were ``discarded" or ``under investigation" \cite{bul}. 

On the other hand, if we use the estimates of rates from Bahia, then the number of cases of Zika and microcephaly should rise to 400 or 2,400 (considering symptomatic or total number of exposed pregnancies respectively), and the number of total microcephaly cases to 2,520 or 12,600, respectively. Since this is the lowest rate in northeast Brazil, it is apparent that there is need for additional studies that can resolve the discrepancies between French Polynesian, Colombian, and Brazilian rates. 

\subsubsection{CDC report on Colombia Dec 16, 2016}

As this paper was being completed, the CDC released a new report on microcephaly in Colombia \cite{Cuevas2016}. New data is provided for the incidence of microcephaly, Zika and the combination of microcephaly and Zika. The results are inconsistent with those reported during the year by the Colombian Ministry of Health. There is an increase of confirmed Zika and microcephaly cases to 147 until epidemiological week 45 which is much higher than the 58 reported by the government until the same week \cite{bul}. While the government reports did not include overall microcephaly rates, the CDC study states there are 476 cases from weeks 5 through 45, a substantial increase from the previous year's 110. The analysis primarily compares 2016 results with 2015 microcephaly background data, but does not consider the potential increase that might result from changes in methodology and scrutiny as a result of the public health emergency due to the epidemic in Brazil. The study acknowledges the difficulties in definition of microcephaly and their potential impact on the counts. Unfortunately, the report does not address the inconsistencies or methodological differences that are present between their numbers and the previously reported values. Since the study is authored in part by physicians associated to the government health ministry, the reasons for the discrepancies are unclear. Moreover, the focus of the paper is on showing that Zika causes microcephaly, and not on the rate at which it does so per maternal Zika infection. The latter is critical for a comparison with Brazil to determine if Zika could be the cause of the majority of cases there. In particular the analysis reported is on a per capita basis and not related to the number of Zika cases. The increased numbers of microcephaly and Zika and microcephaly cases imply a higher rate of Zika induced microcephaly than previous numbers. Hence, the new study increases the possible range of Zika caused microcephaly rates. However, it is unclear whether they are consistent with the high rates of microcephaly per Zika infected pregnancy in northeast Brazil, or with those of French Polynesia. Moreover, absent a study with similar methods in Brazil, it is unclear how to compare government reports in Brazil with the methods that were applied by this study.

The study provides information that is relevant to other aspects of the discussion of microcephaly cases in Colombia. It includes in its count of cases spontaneous abortions, pregnancy terminations and still births that are linked to microcephaly and reports that all of them constitute 44 of the 476 cases reported. This indicates that intentional pregnancy terminations cannot account for a substantial number of missing microcephaly cases as has been suggested in a number of publications, e.g. \cite{McNeil2016}, and as reported for French Polynesia \cite{Cauchemez2016}; moreover the rate of still births is widely discrepant from the cohort study in Rio \cite{Brasil2016}. 

One of the interesting results that is provided is a count of microcephaly cases by department (subnational political units) in Colombia. It is stated that the rates are higher than previous year numbers in areas that do not have propagating Zika infections, above 2,000 meters in altitude. For example the capital region and largest city, Bogota, has a rate of 5.5 per 10,000 births, compared to the previous year country value of 2.1, an increase by 3.4, for a factor of 2.75 increase. A suggestion is made in the paper that the increase might arise from travel or sexual transmission (presumably from traveling male partners). However, in order for this effect to quantitatively account for the observed increase, a large percentage of fathers and pregnant mothers would have to travel to high Zika prevalence areas of the country and over a substantial fraction of the first trimester of pregnancy. The coastal tourist area of Cartagena 
has a rate of 10 per 10,000 births---an increase above background of $7.9$. We can calculate the fraction of population and the duration of visits residents of Bogota would have to visit Cartagena (or equivalent) for the observed increase in Bogota assuming a consistent rate of transmission for tourists and people who live there. The product of population fraction traveling and duration of their stays would have to be 0.43. So, if 50\% of the pregnant couples of Bogota went to Cartagena they would have to be there for 86\% of their first trimester or all but 12 days. If only the partner went there, the rate would have to be multiplied by the rate of sexual transmission. Such a travel schedule is highly improbable even under normal circumstances and even more so after the of Zika infection outbreak and its location became known. We note that the 2015 rate of microcephaly is not reported by department, nor is the rate of Zika infection of microcephaly cases in 2016. We use the country rate as a reference throughout, though the background rate might have been higher in specific departments.

\begin{figure}[tb]
\centering
\includegraphics[width=12 cm]{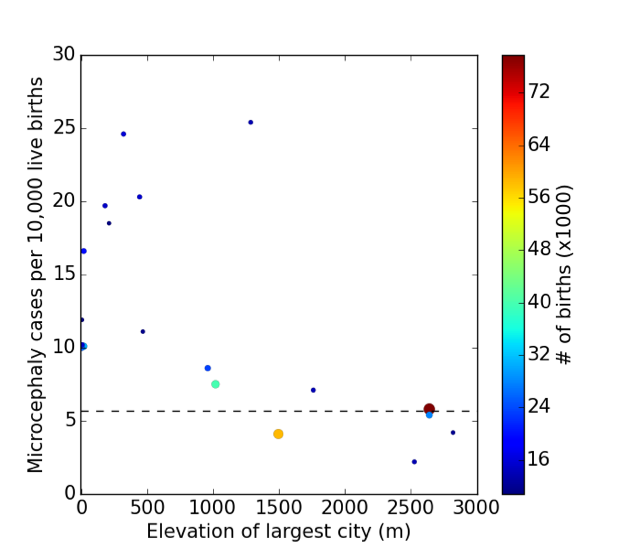}
\caption{Microcephaly versus elevation by Colombian department. Each point represents the reported rate of microcephaly per capita (not per Zika infection) among departments of Colombia versus the elevation of their most populace city. Only departments with more than 10,000 births are shown and color of points reflect the total number of births in the department (scale on right). According to the report \cite{Cuevas2016} high elevation cities (above 2,000 m) do not have self-propagating Zika infections. This includes Bogota and its suburbs which constitute two departments, Bogota and Cundinamarca, at $2,640$ m. The observations suggest a background (non-Zika associated) rate of $5.7$ per $10,000$ births in $2016$ given the study's methodology. The only department consistent with the previous year's background rate of $2.1$ is that of Nariño, whose most populous city Pasto is at $2,527$ m. The rate observed, $2.2$, results from only 3 cases in $14,000$ births, so that statistics are very limited compared to the 60 observed in the Bogota area.}
\label{ColombiaElev}
\end{figure}

It is more reasonable to suggest that the increase in Bogota and other parts of the country at high altitude (especially Cundinamarca, the area surrounding Bogota, together constituting 21\% of all Colombia births), resulted from increased scrutiny and methods that brought background rates close to the median 6.6 prevalence in 17 U.S. states \cite{Cragan2014}. Taking the combined rate in Bogota and Cundinamarca as the reference background would give a 5.7 per 10,000 rate, see Fig. \ref{ColombiaElev}. This would give  283 total background microcephaly cases for the country. Subtracting this from the reported value, 476, yields a number of Zika caused microcephaly cases of 193. This is not far from the rate obtained by taking the 147 cases that are positive for both Zika and Microcephaly  and multiplying by the proportion of 476 untested to 306 tested microcephaly cases, an estimated 228 Zika positive microcephaly cases. This value has to be corrected for background cases that coincidentally have Zika infections due to high prevalence of Zika in the population. The difference between 228 and 193 suggests that about 35 of the cases are due to coincidence which is $35/283= 12\%$ prevalence, or 60,000 infected pregnancies. This is 3 times the number of reported Zika infected pregnancies in the report \cite{Cuevas2016}. Previous estimates have suggested that due to the low rate of reporting of Zika symptoms to Zika infections the rate is 5 times as much, which would increase the coincidence number to 58. Taking this as a range we can summarize the results as suggesting between 160 and 193 Zika caused microcephaly in Colombia. This is about four times the estimates that were obtained from the government reports. 

Such an increase would bring the numbers in Colombia in better agreement with northeast Brazil. Still, the per capita numbers even in the high infection rate regions of Colombia (see figure) do not reach those in northeast Brazil. Absent a comparison that includes Zika infection rates, the CDC study provides new evidence that brings the rate in Colombia closer to those in Brazil but additional information is needed to demonstrate that it is indeed the primary cause. Discrepancies between different measures and positive / negative test results must still be reconciled.


Finally, we note that the CDC report makes several other hard to understand suggestions in their discussion: (1) In considering the number of microcephaly cases that are reported, the paper suggests that the numbers should be considered in relation to a possible reduction in birthrate due to recommendation to delay pregnancy by authorities. However, the reduction in births from 2015 to 2016 is 18,000 or 3.5\%, which would only have an impact on overall microcephaly numbers by such a percentage, about 4 cases; (2) The paper tries to explain the relative increase of number of cases in Brazil and Colombia compared to background multiplicatively (i.e. the ratio of incidence per 10,000 is different both for background observation and for 2016 observations leading to different ratios). However, the effect of Zika on microcephaly rates should be additive (after multiplying by the rate of Zika infections) rather than multiplicative.

To determine the rate of microcephaly per infected pregnancy the number of Zika infected pregnancies must be identified. How many of the microcephaly cases were originally among those identified as Zika infected pregnancies is not reported in the study. Prior studies by the same group reported four cases of microcephaly that were not originally identified as Zika infected pregnancies \cite{ColombiaNEJM}. If we consider the ratio of total and symptomatic cases to be 5:1, and assume 1 in 3 of pregnancies were exposed in the first trimester, and that all pregnancies have terminated, then the number of additional reported microcephaly cases over background is $366$ in $33,000$ or $1.1\%$. Perhaps coincidentally, this is consistent with the reported rate of microcephaly in French Polynesia. Among the significant differences between the studies were the large proportion of abortions relative to live births in the former, which was not present in Colombia. Were we to ignore the asymptomatic cases, then the rate of microcephaly to Zika infection in Colombia would be five fold the number of cases in French Polynesia, again raising the question as to why other cases were not observed there. If we restrict the number of microcephaly cases to those with positive tests for Zika, allowing the rest to be increased reporting due to enhanced scrutiny, we have a rate of $0.44\%$ per first trimester infection. 

More importantly for our discussion, the improvement of agreement between Colombia and French Polynesia, and the increase in number of Zika confirmed microcephaly cases from approximately $50$ to $147$, reduces but does not eliminate the discrepancy between the rate of microcephaly between Colombia and northeast Brazil and among different states of Brazil. Indeed, the reported confirmation of about $50\%$ of tested cases of microcephaly as having evidence of Zika infection stands in contrast to the approximately $15\%$ reported in July in Brazil (see Table \ref{Brazilratio})---consistent with the suggestion that Zika is a cause of only a small fraction of the total cases. We note that if there is another cause of microcephaly in Brazil, even the cases that have evidence of Zika infections would include coincidental co-occurence rather than a necessarily causal relationship, which is not considered in the CDC or other reports.  

In summary, without clarity about differences of methodological aspects of the studies, it is hard to determine what is actually known about Zika and microcephaly in Colombia, leading to highly uncertain conclusions. Increased uncertainty should not be taken to be evidence for Zika as a cause, but rather a need for additional studies that can identify what is the cause. The question is not whether Zika is a cause (which it is, at some rate), but whether it is the cause of the large number of cases in Brazil. A conclusive answer is not yet available and inconsistencies are unlikely to be resolved until this question is directly addressed. 

\subsubsection{Summary of evidence on Zika and microcephaly}

Since the high incidence of microcephaly in Brazil and the spread of Zika virus to other countries, the international community has been waiting to see if an accompanying spread of microcephaly would also occur. After months of uncertainty, the government published data strongly indicates that Colombia will not be seeing a comparably large number of Zika-related birth defects. A new CDC report just released on Colombia implies a larger rate of microcephaly in Colombia due to Zika infections than previously reported, but does not demonstrate consistency with the rate in Brazil due to the absence of reliable Zika infection numbers. Either definite discrepancy or persistent uncertainty suggests a need to reexamine conditions in Brazil, particularly in the northeastern states which saw the majority of microcephaly cases. If Zika alone is not enough to cause large numbers of birth defects, some other factor or factors unique to Brazil are present. Recently several reports have suggested that co-factors are responsible \cite{naturebutler, McNeil2016, Phillips2016}. A cofactor would be one in which Zika is the cause, but the presence of some other environmentally present substance increases the susceptibility. It is also possible that even without Zika infections another cause is responsible if its presence began at about the same time. One possibility is the pesticide pyriproxyfen, which has been largely dismissed as a potential cause or cofactor despite being insufficiently studied \cite{Parens2016}. While authorities claim toxicology studies are sufficient to rule it out as a cause, they surely would not rule it out as a cofactor. Thus the absense of engaging in the possibility of its contribution to the microcephaly by health authorities is inconsistent with the scientific evidence. 


While the total number of microcephaly cases remains low in Colombia and other countries, cases in Brazil continue to rise at the rate of 100 affected births per month (see Fig. \ref{totals}). Since pyriproxyfen may be playing a role in Brazil's disproportionate increase of birth defects, rapid policy action is needed to replace its use as a pesticide until its effects can be more thoroughly studied.

\section{Recommended further studies}

From our research and analysis, we have found many gaps in the available scientific and public health literature. Thus, we suggest areas of study that would contribute to our understanding of the complex public health issues associated with pyriproxyfen and Zika as causes of microcephaly:

\begin{itemize}
\item{Microcephaly definition and counts. One of the central challenges of studies and country data is the definition of microcephaly and the counting of its cases. In order to achieve consistency among studies it is necessary to go beyond considering a single definition and number. What is needed is a reporting of how the number of cases changes as the threshold cranial circumference is varied. This distribution is a much more robust measure than a single number.}
\item{Analysis of the geographic and time dependence of the use of pyriproxyfen in Brazil in relation to microcephaly occurrence.} 
\item{Observations of the concentration of pyriproxyfen and its breakdown products in containers of drinking water, and in tapped water, over time after administration.} 
\item{Animal studies of pyriproxyfen and its metabolites. The existing studies upon which the approval of pyrirproxyfen are based are statistically weak and assumptions used are not well justified. Moreover, additional studies should be done on breakdown products of pyriproxyfen in water, with and without sunlight, which may have distinct effects.} 
\item{Retinoid X receptor binding/ligand tests for pyriproxyfen and its biological and environmental metabolites.}
\item{A more systematic understanding of the effects of juvenile hormone analogs and their toxicity across mammalian models, including the effects of genetic variation and multiple potential mechanisms of action (retionid X receptor, thyroid mechanisms, etc).} 
\item{Tests of reliability of traditional toxicological test assumptions for hormones and other regulatory molecules.}
\end{itemize}



\section{Conclusion}
This paper analyzes the potential causal connection between the pesticide pyriproxyfen and microcephaly, as an alternative to Zika. Pyriproxyfen is a juvenile hormone analog, which has been shown to be cross reactive with retinoic acid, part of the mammalian regulatory system for neurological development, whose application during development causes microcephaly. This causal chain provides ample justification for pursuing a careful research effort on the role of pyriproxyfen in neurodevelopmental disorders. Counter to stated claims, existing studies of neurodevelopmental toxicity by Sumitomo, its manufacturer, provide some supportive evidence for neurodevelopmental toxicity including low brain weight in rat pups. The large-scale use of pyriproxyfen in Brazil and its coincidental timing with an increase in microcephaly cases should provide additional motivation. We believe that this evidence is strong enough to warrant an immediate cessation of pyriproxyfen application to Brazilian water supplies until additional research can be carried out on its neurodevelopmental toxicity. Alternative hypotheses about causes or factors affecting the incidence of microcephaly should be considered \cite{naturebutler, McNeil2016, Phillips2016}. Where insecticides are considered essential, Bt toxin is considered a safe alternative that has been claimed to be used in water systems in Recife since 2002 \cite{BtToxin}. Other vector control methods ranging from better sealed water containers, to ovitraps may also be used \cite{Regis2013,Regis2008,Bar-Yam2016}. 



\textbf{Acknowledgements:}
We thank Dan Evans and Audi Byrne for helpful discussions, and Keisuke Ozaki for helpful communications about Sumitomo toxicology tests.


\begin{thebibliography}{20}
\bibitem{error} There is a discrepancy between the number of reported (42) and the number of cases shown in Fig. 1 of that paper (41), the numbers we cite are those present in the figure as it is unclear which week the additional exposure is in.
\end{thebibliography}
\end{document}